\pdfoutput=1
\documentclass[sigconf, nonacm]{acmart}

\usepackage{cleveref}
\usepackage{listings}
\usepackage{xcolor}

\lstdefinestyle{python}{
  language=Python,
  morekeywords={as, True, False, None},
  deletekeywords={from} % used in fig:sir-example, but nowhere else
}

\begin{document}
\title{Vidformer: Drop-in Declarative Optimization for Rendering Video-Native Query Results}

\author{Dominik Winecki}
\orcid{0009-0009-8632-3102}
\affiliation{
  \institution{The Ohio State University}
  \city{Columbus, Ohio}
  \country{USA}
}
\email{winecki.1@osu.edu}

\author{Arnab Nandi}
\orcid{0000-0002-4138-603X}
\affiliation{
  \institution{The Ohio State University}
  \city{Columbus, Ohio}
  \country{USA}
}
\email{nandi.9@osu.edu}

\begin{abstract}
When interactively exploring video data, \textit{video-native} querying involves consuming query results as \textit{videos}, including steps such as compilation of extracted video clips or data overlays.
These video-native queries are bottlenecked by \textit{rendering}, not the execution of the underlying queries.
This rendering is currently performed using post-processing scripts that are often slow.
This step poses a critical point of friction in interactive video data workloads: even short clips contain thousands of high-definition frames; conventional \texttt{OpenCV}/Python scripts must \emph{decode $\rightarrow$ transform $\rightarrow$ encode} the entire data stream before a single pixel appears, leaving users waiting for many seconds, minutes, or hours.

To address these issues, we present \textsc{Vidformer}, a drop-in rendering accelerator for video-native querying which,
(i) transparently \emph{lifts} existing visualization code into a declarative representation,
(ii) transparently optimizes and parallelizes rendering, and
(iii) instantly serves videos through a Video on Demand protocol with just-in-time segment rendering.
We demonstrate that Vidformer cuts full-render time by \mbox{2--3$\times$} across diverse annotation workloads, and, more critically, drops time-to-playback to 0.25--0.5\,s.
This represents a \mbox{400$\times$} improvement that decouples clip length from first-frame playback latency, and unlocks the ability to perform interactive video-native querying with sub-second latencies.
Furthermore, we show how our approach enables interactive video-native LLM-based conversational querying as well.
\end{abstract}


\maketitle

\section{Introduction}\label{sec:intro}

\begin{figure}
    \centering
    \includegraphics[width=\linewidth]{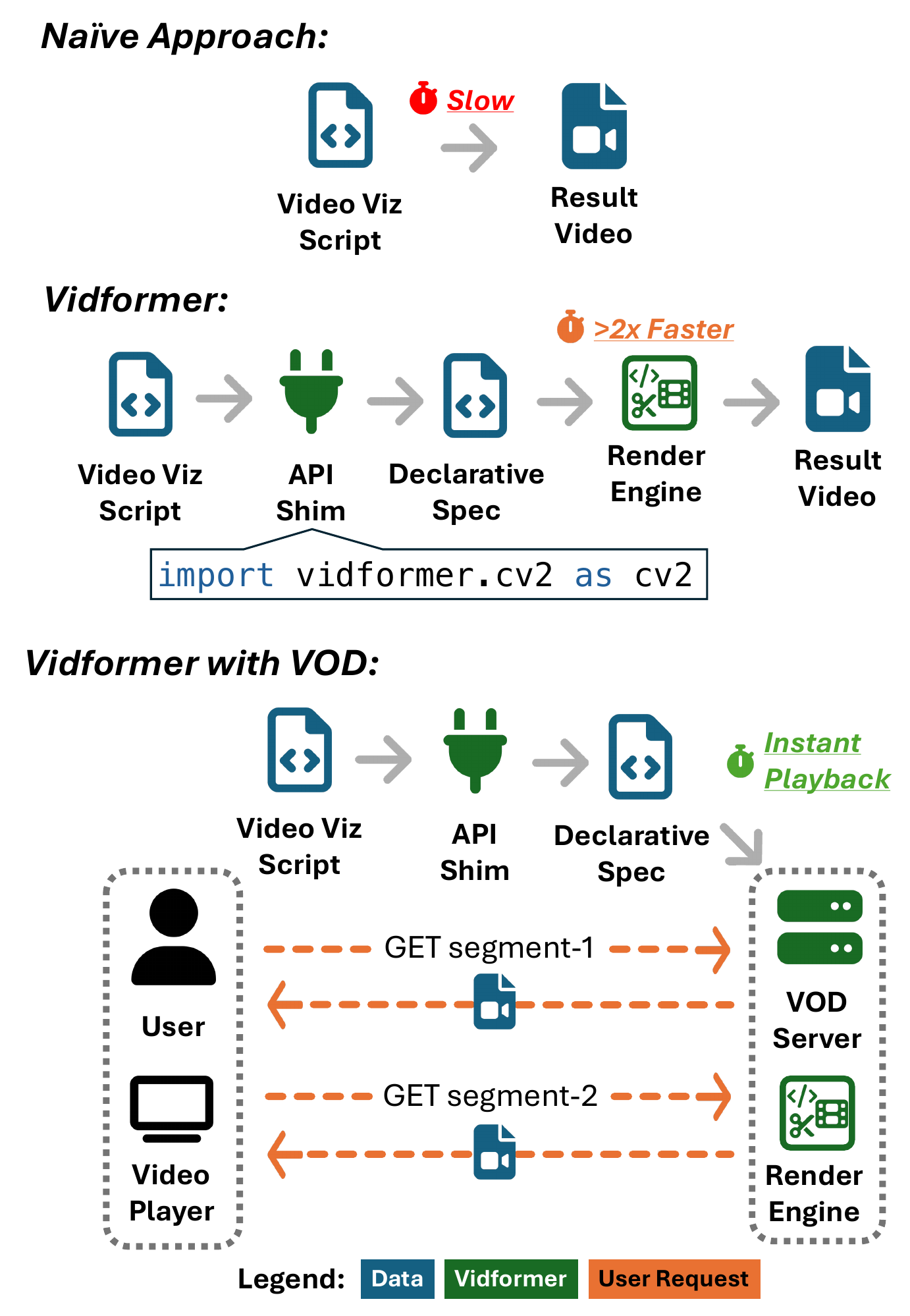}
    \vspace{-5mm}
    \caption{The naïve approach for rendering visualization videos is exceedingly slow. \textit{Vidformer} uses a drop-in API shim to convert imperative code into declarative specifications and efficiently accelerate result rendering. Vidformer's Video On Demand server then materializes segments just-in-time.}
    \label{fig:vf-methods}
    \vspace{-5mm}
\end{figure}

Video data has become ubiquitous alongside tremendous advancements in computer vision and multimodal data processing in recent years.
Domain-specific computer vision models now outperform humans using modest computational resources, while modern large language models (LLMs) and vision-language models (VLMs) are inherently multimodal, enabling zero-shot interactions with text, images, audio, and video.
Additionally, video database management systems (VDBMSs) have made significant strides in running models over videos~\cite{kang_viva_2022,kakkar_eva_2023}, largely solving inference at scale.
These advancements have made limitations elsewhere in video data pipelines evident.
We focus on one significant point of friction: \textit{rendering video/data visualization videos.}

In computer vision, annotating videos with model results is a critical technique for debugging and understanding models.
Bounding boxes and object masks provide an intuitive understanding of the strengths and limitations of detection and segmentation models.
Additionally, attention heatmaps allow visualizing the impact of individual features.
Similarly, video databases are adept at answering queries over video datasets, such as which videos or frames match a search, and return relational/tabular results that can be used to create result compilation videos.
Numerous other applications create similar video data visualizations, such as overlaying relevant telemetry information onto drone footage or synchronizing manufacturing sensor data with video streams from assembly lines.
These use cases all share a similar goal: combining video streams with conventional data about them to create a new, enhanced visualization video.
Such video data visualization workflows are widely used in scientific, commercial, and industrial applications.

We have found that these varied tasks have nearly identical implementations: a script is written in a general-purpose programming language to perform the edit, creating a transformed visualization video.
This approach offers unmatched interoperability with conventional data processing tools, along with the unbounded expressivity of Turing-complete scripts.
However, this approach to rendering visualization videos is slow; even fast visualization scripts take many seconds or minutes to export a video.
This significantly slows down iteration speed and incentivizes users to minimize their use of visualization videos and, when they do use them, to create only short videos.
This poor experience is primarily a matter of tooling.
One can run a conventional video editor, such as Adobe Premiere Pro or Apple Final Cut, on commodity hardware and watch edited videos in real time, demonstrating a clear tooling gap.
In this paper, we detail a novel solution to \textit{instantly} present script-defined video data visualizations to users.

Accelerating the rendering of video data visualizations requires resolving the conflict between expressivity and the complexity of control flow.
Simple tasks can be hard-coded into an interface to achieve reasonable performance, such as baking overlays into videos when they are first encoded.
However, this approach cannot adapt to more complex tasks, as implementation complexity scales exponentially with the addition of functionality.
Bespoke video data visualization scripts, such as those shown in \Cref{fig:cv2-example}, have excellent expressivity and interoperability with data, computer vision, and multimedia ecosystems.
However, they fully render output videos and tightly integrate imperative control flow with unbounded complexity, precluding most opportunities for transparent optimization.
Similarly, while it is possible to incorporate efficient algorithms into these scripts, this significantly increases their complexity.
Even perfectly optimized, hardware-accelerated visualization scripts of this sort cannot reach the required speeds, as they fully render the result videos.

Addressing these challenges is essential as video data workflows continue to proliferate.
Video has emerged as a prominent data type across numerous scientific, commercial, and consumer applications.
Despite this growth in video data volume, video has not experienced the same level of innovation in interactive interfaces as other data types, resulting in video data being slow to explore and analyze.
Displaying video data visualizations quickly is a crucial step toward bridging this gap.
In addition to the immediate benefit of accelerating existing workloads, a second-order effect is enabling new interaction paradigms.
A long-standing chicken-and-egg problem is that novel features and applications are not developed because they would be too slow for practical use, and there is little incentive to improve their speed, as few features or applications exist to necessitate performance improvements.
Solving the performance challenges of video data visualization allows novel experiences to be built on top of them.
Namely, LLM-driven video-native querying, which we detail in \Cref{sec:llm-querying}.

This paper presents \textsc{Vidformer}, a novel approach for drop-in acceleration of video data visualization scripts.
Our approach converts imperative video data visualization code into a declarative specification, then efficiently optimizes, parallelizes, and executes it to serve a transformed visualization video.
We begin by detailing our method for lifting existing code into a declarative representation via symbolic computation (\Cref{sec:lifting}).
We implement an API compatibility layer for the most popular of these libraries, OpenCV's Python library, enabling a one-line \texttt{import vidformer.cv2 as cv2} patch to convert existing code.
We then detail the design of our rendering engine (\Cref{sec:execution}), which executes declarative transformations on video, as DBMSs execute SQL to transform relational data.
Finally, we propose an approach to use Video on Demand (VOD) streaming protocols to serve deferred video results (\Cref{sec:vrod}).
Short segments are lazily materialized just-in-time.

From a user's perspective, videos are instantly available; users can play and skip around the video, just as they did before.
Most critically, we focus on creating a versatile solution; for our system to be effective, it does not just need to work in the \textit{typical} case; it needs to work in \textit{every} case, so we focus on worst-case video access patterns.
By designing generalized solutions, our system is as expressive as general-purpose scripting languages at video annotation, transformation, and visualization, even on the most complex tasks.

Vidformer treats video transformation and rendering as a data-processing task, transparently augmenting the rendering control flow to optimize and defer the materialization of visualization videos while maintaining the full expressivity of existing scripting languages.
Our evaluation shows declarative lifting and our render engine speeds up rendering by $2-3\times$, and streaming results over $400\times$ faster, delivering sub-second time-to-playback to users.

\section{Related Work}\label{sec:relwork}

\subsubsection*{Visualizing Computer Vision Model Outputs}
We have found that Supervision~\cite{Roboflow_Supervision} is the predominant higher-level tool for annotating images and videos with computer vision results.
Model results, primarily from YOLO~\cite{7780460} and SAM~\cite{Kirillov_2023_ICCV} derivatives, can be loaded into a common detection format that stores bounding boxes, masks, classifications, confidence scores, and object tracking IDs.
Different annotators can then be used to visualize detections on their underlying frames.
Supervision is built on top of OpenCV internally, and we use Supervision's annotators in our evaluation.
In our survey, virtually \textit{all} computer vision visualizations used Supervision or OpenCV, with the sole exception being PyTorch's~\cite{paszke2019pytorchimperativestylehighperformance} \texttt{transforms} module.
This ubiquity is critical: LLM training datasets are packed with code using OpenCV and Supervision, making LLMs proficient users of these libraries.

\subsubsection*{Video Databases That Return Videos}
Many video databases can return video clips~\cite{hwang_querying_1996,arslan_semantic_2002,catarci_bilvideo_2003,kang_challenges_2019,zhang_fast_2019,zhang_equi-vocal_2023,kang_viva_2022,kakkar_eva_2023,kittivorawong_spatialyze_2023}, often clipped segments, but only return \textit{pointers} to clipped time-spans, not actual videos.
ApertureDB~\cite{aperturedb} has a dedicated command for extracting clips from videos.
Additionally, a few systems provide limited support for annotating videos, but this is almost exclusively done with ad hoc, standalone OpenCV~\cite{opencv_library} scripts.
For example, EVA~\cite{kakkar_eva_2023} and Spatialyze~\cite{kittivorawong_spatialyze_2023} both include Python scripts that use OpenCV to draw bounding boxes.

\subsubsection*{Declarative Video Editing}
Incorporating data into video editing has been explored for annotation of time-series video streams~\cite{mackay_virtual_1989,mackay_eva_1989,fouse_chronoviz_2011,mackay_diva_1998}, with Mackay \& Davenport stating that ``video becomes an information stream, a data type that can be tagged and edited, analyzed and annotated''~\cite{mackay_virtual_1989}.
Notably, DIVA~\cite{mackay_diva_1998} offers a ``stream algebra'' for annotating and rearranging video stream segments, and to the best of our knowledge, it is the first declarative video editor (DVE) system.
Outside data annotation tasks, DVE and script-based editing have been explored for conventional multimedia editing~\cite{andersen_super_2017,matthews_videoscheme_1993}.
With the advent of the web, numerous media platforms emerged that utilized DVE concepts.
Orcc~\cite{yviquel_orcc_2013} provides graphical dataflow-based editing.
NUBOMEDIA~\cite{garcia_nubomedia_2017} is a WebRTC-based data plane for media applications, and Hop~\cite{serrano_programming_2007} is a unified language and web framework for this.
Similarly, Videostrates~\cite{klokmose_videostrates_2019} is a collaborative video editing toolkit with a declarative HTML and CSS interface to transform media. Our prior work, V2V~\cite{winecki2024_v2v}, articulates a vision for a DVE purpose-built for creating video results from queries.
It models video editing as a composition of filters over frames, and videos are sequences of (timestamp $\times$ frame expression) pairs represented in a domain-specific language (DSL).
Vidformer builds on this vision, expanding expressivity beyond the limitations of a DSL while reducing rendering times to usable levels.

In open source, \textit{editly}~\cite{finstad_mifieditly_2024} is a JSON-based DSL for editing videos, while revideo~\cite{noauthor_redotvideorevideo_2024}, an open-source alternative to remotion~\cite{noauthor_remotion_nodate}, has a Typescript/React interface.
In industry, companies are offering declarative video editing for content creation~\cite{noauthor_shotstack_nodate,noauthor_remotion_nodate,noauthor_creatomate_nodate}, all through APIs that take JSON specifications and return a video.
VideoDB~\cite{noauthor_videodb_nodate} supports streaming video results via HLS for creating compilations, which are hard-coded into its API.
Each of these is designed for conventional video editing tasks and lacks the expressivity needed for even simple annotation.
\paragraph{The need for a drop-in approach:}
Declarative video editing has been studied intermittently for decades.
Despite tremendous growth in the volume and applications of video data, few systems exist or are in use.
We believe this is due to a focus on creating declarative DSLs, which are difficult to adopt.
For conventional video editing tasks, such DSLs are not competitive with the interface of graphical editors (e.g., Adobe Premiere Pro, Apple Final Cut), nor are they expressive enough to be useful for programmatic tasks like data annotation. In this paper, we build on a drop-in, declarative internal model for its performance benefits, but hide this from users, in much the same way as existing graphical video editors. This approach is analogous to Modin~\cite{petersohn_towards_2020}, which positions itself as a drop-in replacement for scaling Pandas code.
Vidformer scales and optimizes video pipelines using a portfolio of techniques discussed in this paper, while abstracting away implementation details, allowing users to adopt database techniques friction-free.

\subsubsection*{Automatic Video Editing}
Automatically editing videos is an established problem~\cite{bocconi_semantic-aware_2004}.
Many of these are the common \textit{supercut}-style edits~\cite{andersen_super_2017}, montages of short clips.
In addition, there has been a recent surge in automatic video editing directly using machine learning~\cite{zhang_ai_2022}, such as automatic editing style transfer~\cite{frey_automatic_2021}.
This demonstrates interest in compiling useful videos in response to specific user demands across a wide range of fields.
However, these focus solely on how to logically edit videos, not on how to materialize a physically rendered result.

\subsubsection*{Symbolic Data Processing Libraries}
Symbolic data processing is a foundational technique for compute-intensive operations.
Machine learning libraries like Tensorflow~\cite{tensorflow2015-whitepaper} and PyTorch~\cite{paszke2019pytorchimperativestylehighperformance} use symbolic operations on tensors, which enable transparent gradient calculation and acceleration via GPUs/TPUs.
Similarly, Apache Spark~\cite{10.1145/2934664} allows symbolic transformations on datasets to execute across compute clusters.
While these systems utilize bespoke interfaces that require some user effort, newer approaches leverage API compatibility to accelerate tasks with minimal user input.
Modin~\cite{petersohn_towards_2020} is a drop-in dataframe processing accelerator supporting multiple backends.
It uses an API compatibility layer, making adoption as simple as a one-line \texttt{import modin.pandas as pandas} in Python.
CuPy~\cite{cupy_learningsys2017} does the same for accelerating NumPy/SciPy code using GPUs.
Similar to these prior works, we utilize a Python one-line import API shim with symbolic n-dimensional array references to declaratively model operations/transformations.

\subsubsection*{Progressive and Online Results}
Instant result consumption has long been valued in \textit{non-video} analytical query settings, motivating
\emph{online aggregation} techniques that return statistically valid partial answers
while a query scans the data~\cite{hellerstein_online_1997,haas_ripple_1999,10.1145/1412331.1412335}.
Consequently, achieving \emph{truly interactive} video-native queries demands new operator pipelines, optimizations, and lazy materialization strategies such as those we propose in this work.

\section{Problem Definition}

Our objective is to transparently accelerate the rendering of video data visualization scripts.
Such scripts use code written in a general-purpose programming language, namely Python, to create a new video by combining existing video and additional data.
For example, drawing bounding boxes on a video to understand a computer vision model, or creating a compilation of video segments that match a search query.
Our task is to minimize time-to-playback latency of presenting a user with the \textit{output} of such scripts.

While applications are broad, they share some additional operating environment constraints that must be considered:

\begin{itemize}
    \item \textbf{Correctness:} Regardless of complexity, output must be pixel-for-pixel identical to the existing methods.
    \item \textbf{Embedded:} Video data visualizations exist within larger interfaces. For example, computer vision work typically occurs in interactive notebooks, while video search results are integrated within larger web applications.
    \item \textbf{Over-the-network:} Applications typically operate on remote systems served over (potentially slow) networks.
\end{itemize}

These constraints ensure that the resulting system is practically usable in existing applications.

\section{Declarative Lifting of Video Data Visualization Scripts}\label{sec:lifting}

\begin{figure}
     \begin{lstlisting}[style=python]
import cv2             # OpenCV
import pandas as pd    # Dataframe library

df = pd.read_csv("objects.csv")
cap = cv2.VideoCapture("in.mp4")
writer = cv2.VideoWriter("out.mp4", ...)

i = 0
while True:
    ret, frame = cap.read()
    if not ret:
        break

    cv2.putText(frame, f"This is frame {i}", ...)
    for _, row in df[df.frame == i].iterrows():
        x1, y1, x2, y2 = row["xyxy"]
        cv2.rectangle(frame, (x1, y1), (x2, y2), ...)
    
    writer.write(frame)
    i += 1

cap.release()
writer.release()
\end{lstlisting}
    \caption{A typical video data visualization script written in Python using OpenCV. Each frame is annotated with its number and bounding boxes from an annotation file are drawn on each frame.}
    \label{fig:cv2-example}
\end{figure}

We introduce our approach to lift existing imperative video transformation code into a declarative specification.
We use the Python OpenCV library for demonstration, as it is widely used, with 26 million downloads per month.
A search on GitHub found 3.9 million files using it, 99.8 thousand of which create a \texttt{VideoWriter} object.
This suggests that roughly $2.6\%$ of OpenCV users are performing some video data visualization, annotation, or editing task.
An example of such a script is shown in \Cref{fig:cv2-example}.

\begin{figure}
    \centering
\begin{lstlisting}[style=python, escapeinside={(*}{*)}]
vidformer.pixfmt(
    opencv.rectangle(
        opencv.putText(
            vidformer.pixfmt(
                (*\textit{\textcolor{gray}{$\langle$Frame 0 from in.mp4$\rangle$}}*),
                from="yuv420p",
                to="bgr24"),
            "This is frame 0",
            ...),
        (10, 30),
        (200, 50),
        ...),
    from="bgr24",
    to="yuv420p")
\end{lstlisting}
    \caption{Internal representation of the declarative specification for the first output frame from the script in \Cref{fig:cv2-example} when the single-line \texttt{import vidformer.cv2 as cv2} edit is made. OpenCV functions are composed with constant arguments, and underlying frames are referenced declaratively.}
    \label{fig:sir-example}
\end{figure}

\subsection{Declarative Data Model}

We represent video transformations by storing the construction of each \textit{output frame}.
Each output frame is represented as a \textit{frame expression}, a composition of filter functions, constant data values, and input frame references.
For example, the first frame produced by the script in \Cref{fig:cv2-example} is shown in \Cref{fig:sir-example}.
Here, ``\texttt{opencv.putText}'', ``\texttt{opencv.rectangle}'', ``\texttt{vidformer.pixfmt}'' are filter functions,\linebreak \textcolor{gray}{$\langle$Frame 0 from in.mp4$\rangle$} is an input frame reference to the first frame in \texttt{"in.mp4"}, and the remaining strings, integers, and tuples are constant data values.
We store one frame expression for each output frame as an abstract syntax tree (AST).
Since frame expressions are deeply nested, verbose, and repetitive, we store sequences of frame expressions in a flattened AST data structure with interning to minimize memory usage.

We also apply type semantics to frames to quickly catch errors.
A frame's type combines its resolution and pixel format; for example, in \Cref{fig:sir-example}, \textcolor{gray}{$\langle$Frame 0 from in.mp4$\rangle$} has type \texttt{<1280x720, yuv420p>}.
This is the encoded format of the \texttt{"in.mp4"} video file.
Then a function converts it to \texttt{<1280x720, bgr24>} to apply OpenCV filters, and it is then converted back to \texttt{yuv420p} for final encoding.
We have found that many systems, including OpenCV, default to \texttt{rgb24} or \texttt{bgr24} formats, despite most videos being in a \texttt{yuv*} format.
This conversion is not always required, leading to unnecessary pixel format conversions and increased memory usage.
By allowing different in-memory pixel formats in filters, we enable more efficient memory usage and rendering.

\subsection{Lifting via Symbolic Computation}

We propose an OpenCV API shim with symbolic references to frames and filters to compose declarative specifications.
Conceptually, rather than representing frames as n-dimensional arrays, we introduce a \texttt{Frame} object, which mimics a virtual n-dimensional array and tracks transformations to construct a frame expression.
\texttt{cv2.VideoCapture} objects provide an iterator over frames in video sources, OpenCV functions, such as \texttt{cv2.putText}, map to filters in our data model, and \texttt{cv2.VideoWriter} objects collect rendered frames to create specifications.
Since no video frames are decoded, transformed, or encoded, the symbolic code executes many orders of magnitude faster and with a fraction of the memory usage compared to the non-symbolic equivalent.

When a \texttt{cv2.VideoCapture} object is instantiated, the source video's resolution, pixel format, and number of frames are retrieved.
Successive \texttt{read()} calls create and return \texttt{Frame} objects storing their resolution, pixel format, and construction.
The construction is a function composition of filter function calls needed to construct the frame.
Frame objects maintain a user-facing NumPy-compatible interface.
We apply an extra optimization for pixel formats.
Most videos are stored in \texttt{yuv420p} or similar pixel formats, and by default OpenCV converts all frames to \texttt{bgr24} when decoding.
Therefore, to prevent unnecessary pixel format conversions, \texttt{Frame} objects present in their equivalent OpenCV format but internally maintain a true format that remains in the input video's native format until a filter requiring a pixel format conversion is applied.

Each supported OpenCV function is implemented with the same function signature and updates the \texttt{Frame} objects.
There is not an exact one-to-one mapping between OpenCV functions and Vidformer-compatible OpenCV filter functions; many OpenCV/\texttt{cv2} functions mutate frames in place, whereas Vidformer filters are purely functional and always return a single frame; our API compatibility layer hides these differences.
A few utility functions that lie outside our conceptual data model, such as \texttt{cv2.getTextSize()}, are passed to the actual \texttt{cv2} library.
Transformed \texttt{Frame}s are written to a \texttt{cv2.VideoWriter}.
These are converted into their declarative representation and appended to the video's specification.
Once this writer is closed, the video specification is complete.

\subsubsection{Beyond Python/OpenCV Scripts}
We primarily focus on Python scripts using OpenCV, as this combination is ubiquitous in practice, but our approach generalizes beyond these.
We have validated that other frontends can be supported using either symbolic computation (e.g., with FFmpeg filters) or transpilation (e.g., with DVEs), including every system mentioned in \Cref{sec:relwork}.
Many higher-level tools, such as Supervision~\cite{Roboflow_Supervision}, are built on top of OpenCV and provide better ergonomics for specific tasks.
Our approach can support these with minimal effort; our implementation supports Supervision with \texttt{import vidformer.supervision as sv}, which is only necessary to remove a single \texttt{assert frame is numpy.ndarray} check.

\subsection{Masks, Heatmaps, and other Raster Data}\label{sec:raster-data}

Some visualization types incorporate raster data, such as object masks or heatmaps, which can pose challenges when naively embedded as a constant parameter in a specification.
To address this, we treat these n-dimensional data arrays as frames in a lossless video stream, replacing these data with newly created \texttt{Frame} references.
When visualizing the results of a segmenting object detector (e.g., Segment Anything~\cite{Kirillov_2023_ICCV}), we encode a new video packing each object mask into a single frame with a single-channel pixel format, namely \texttt{gray8}, and compress the stream with the lightweight and lossless FFV1~\cite{rfc9043} codec.
This enables fast space-efficient compression, and, despite the unconventional use of object masks as video frames, access remains fast due to the design of our rendering engine's decoding system (\Cref{sec:ooo}).

\section{Declarative Rendering of Result Videos}\label{sec:execution}

\begin{figure}
    \centering
    \includegraphics[width=0.9\linewidth]{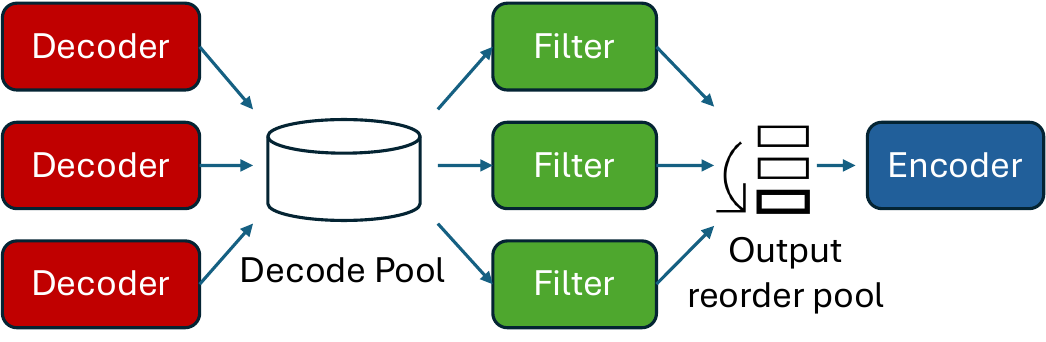}
    \vspace{-1mm}
    \caption{Rendering engine architecture diagram.}
    \label{fig:thread-arch}
\end{figure}

We detail our declarative video transformation rendering engine, which runs declarative specifications to render videos.
We take a greedy, coordinated-multiprocessing approach, as shown in \Cref{fig:thread-arch}.
A group of decoder threads decodes the required frames and places them in a decoded frame pool.
A group of filter threads then renders output frames using decoded input frames from the decode pool.
Filter threads insert rendered output frames into an output reorder pool.
The output reorder pool serves as a temporary buffer to ensure that frames are sent to the encoder in the correct order.
The encoder pulls frames from the output reorder pool and writes the resulting video to disk.
Each output frame is created from a frame expression, which is a dependency tree of composed filter functions.
This makes the filter and encoding components relatively simple; execution can be parallelized regardless of the complexity of frame expressions or the extent to which they vary between successive output frames.
However, filters depend on decoded input frames, and ensuring frames are efficiently loaded in time for all access patterns poses a significant scheduling challenge.
We refer to this as the out-of-order frame access problem.

\subsection{Out-of-Order Frame Access}\label{sec:ooo}

To reach parity with existing methods, supporting \textit{all} video data visualization/transformation tasks is essential.
Most use frames in sequential order, but others do not, such as sampling one frame per second, only using keyframes, reversing a video, using frames from multiple videos, or displaying multiple frames side-by-side.
Additionally, when rendering many output frames in parallel, even sequential frame access patterns become erratic.
Further, as detailed in \Cref{sec:raster-data}, data is compressed in videos as well; for example, object masks from computer vision segmentation models are stored as individual frames in a losslessly compressed video.
Rendering a frame that highlights object masks requires decoding all object-mask frames, which may include dozens of objects per frame scattered across the object-mask video stream.
Such arbitrary frame access patterns conflict with how videos are encoded and stored.
Video codecs group frames into Groups of Pictures (GOPs) to exploit temporal locality in video.
GOPs start with a keyframe/I-frame, followed by dependent P (progressive) and B (bi-directional) frames that reference artifacts from previously decoded frames within the GOP.
Smaller GOP sizes enable efficient random access, and All-I (every frame is a keyframe) compression is preferred in conventional video editing workflows.
Larger GOP sizes allow higher compression ratios, but this comes at the expense of higher memory usage during decoding and additional compute to seek a new position in the stream.
Sporadic access patterns result in \textit{decode amplification} proportional to GOP size, often on the order of $30-300\times$.
To avoid this detrimental slowdown, we designed our rendering engine to avoid such accesses when possible and to efficiently parallelize them when they are unavoidable.

\subsection{Rendering Engine for Result Videos}

\begin{figure}
    \centering
    \includegraphics[width=\linewidth]{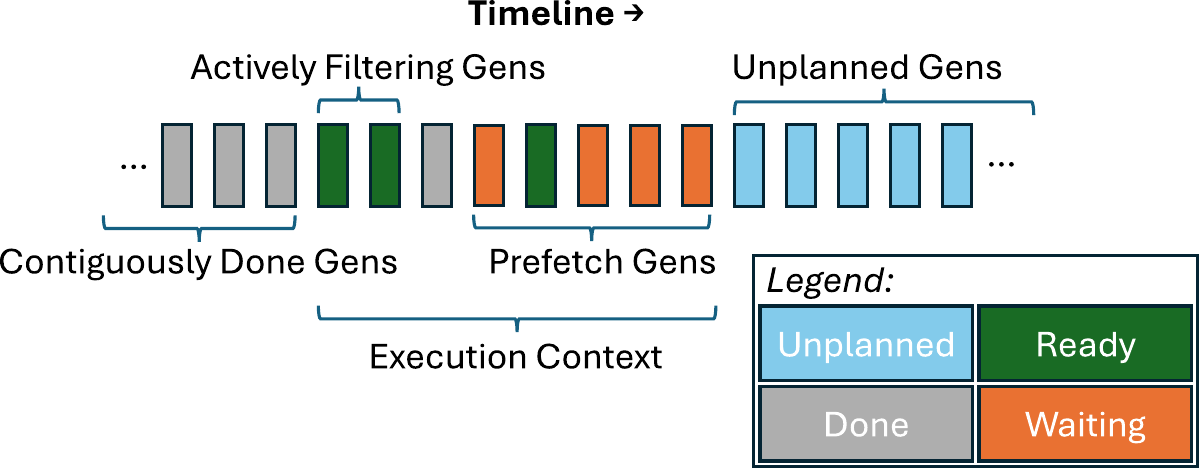}
    \vspace{-5mm}
    \caption{Generation states of the decode pool during execution. Source frames are prefetched prior to filtering.}
    \label{fig:oframe-schedule}
\end{figure}

We design a rendering pipeline to efficiently use memory in rendering, as seen in \Cref{fig:oframe-schedule}.
To better distinguish between input frames and output frames in this paper, we refer to each output frame index as a \textit{generation} or \textit{gen}.
Each generation exists in one of three states: Unplanned, Active, and Done.
Generations are planned in sequence, which adds them to the ActiveGens set.
The scheduler watches this set and instructs the decoders to load all input frames needed by active generations.
A generation is ``Ready'' if all needed input frames are present in the decode pool; otherwise, it is ``Waiting''.
Once an active generation becomes ``Ready,'' the scheduler assigns a filter thread to render the corresponding output frame and insert it into the output reorder pool.
Once the encoder removes the output frame from the output reorder pool, the generation is marked as ``Done'' and removed from ActiveGens.

\subsubsection{Decoders}

Decoding a video segment must begin at the start of a GOP.
Since decoders process GOPs independently, we model a decoder as an iterator of frames from a \textit{single} GOP.
This is the same technique used by Scanner~\cite{poms_scanner_2018} for video batch processing.
We represent each decoder's state as their target GOP and the set of frames yet to be decoded, which we refer to as their FutureSet.
This is a set, not a position, because we expect decoders to output frames in arbitrary order.
For example, a three-frame GOP \(\langle 1,2,3 \rangle\) with types \(\langle I,B,P \rangle\) is stored as \(\langle I,P,B \rangle\) and decoded in order \(\langle 1,3,2 \rangle\).
Existing iterator/stream-based systems~\cite{gstreamer,ffmpeg_developers_ffmpeg_2024} reorder frames to present a continuous stream by using a reorder buffer.
Instead, we immediately place frames into our global decode pool.
This is more memory-efficient than per-decoder reordering, as it globally optimizes memory usage rather than per-video stream, and allows future generations to begin filtering earlier than would otherwise be possible.

\subsubsection{Scheduling and Pool Management}

Our declarative specification provides frame expressions for each \textit{output} frame, and we extract the set of needed \textit{input} frames for each output frame.
We refer to this list of sets of needed frames as the \textit{schedule}.
We then define a NeedSet as all the frames needed by active gens:
$$\text{NeedSet} = \bigcup_{g \in \text{ActiveGens}} \text{schedule}[g]$$
The scheduler plans generations, adding them to ActiveGens until the NeedSet no longer fits in the decode pool or the number of active generations reaches the prefetch window size.
Following this, the scheduler assigns GOPs to decoders to ensure all items of the NeedSet are present in the decode pool.
To do so, the scheduler finds all frames in NeedSet that are not present in the decode pool or FutureSet of any existing decoder.
Of these, it selects the frame needed by the lowest-numbered generation that is not yet ``Done'' and assigns the GOP containing that frame to the decoder.
The scheduler repeats this process until all decoder threads are assigned.

Each decoder thread monitors the NeedSet, making progress whenever a frame in its FutureSet is also in the NeedSet but not yet in the decode pool.
Decoded frames are placed into the decode pool if they are needed by any generation that has not yet been completed.
If no frames from the FutureSet are in the NeedSet but not in the decode pool, then the decoder will stall until this condition changes.
Once a decoder thread finishes decoding its assigned GOP, it waits for a new assignment from the scheduler.
Additionally, we allow decoders to abandon their assigned GOP early if each of these conditions is met:
\begin{itemize}
    \item \textit{The decoder is stalled.}
    \item \textit{There are more critical frames to decode:} There are frames in the NeedSet and not in the decode pool, which are needed by a sooner generation than all frames in the decoder's FutureSet.
    \item \textit{The decoder is the least needed decoder:} The soonest-needed future frame from the current decoder is needed after (or at the same time as) the soonest-needed future frame in all other decoders.
\end{itemize}
This decoder GOP abandonment policy keeps all decoder threads busy and stops the system from deadlocking.

Decoder threads attempt to insert each decoded frame into the frame decode pool, which has a fixed size.
Since the NeedSet can never exceed the pool size, it functions as a reserved portion of the pool, and the remaining portion is used as a cache.
If the frame is in the NeedSet, it forces the insertion, triggering an eviction if necessary.
If the frame is not in the NeedSet, insertion is attempted based on the eviction policy.
We use optimal caching eviction, always evicting the frame needed by the least-soonest incomplete generation:
$$
\begin{array}{l}
     \text{NextNeededGen}(f) = \\
     \begin{cases} 
\min \{g \in \text{NotDoneGens} \mid f \in \text{schedule}[g]\} & \text{if } \exists g, f \in \text{schedule}[g] \\
\infty & \text{otherwise}
\end{cases}
\end{array}
$$

We use the decode frame pool to mask gaps in temporal locality, easing the burden when decoding with out-of-order accesses.
To ensure Vidformer can accelerate even the most complex video transformations, we must be able to utilize the pool as an effective cache, regardless of the frame access pattern.
Filter threads execute on earlier frames in the execution context, and later frames are fetched in advance.
If the scheduler plans too far into the future, it leaves no slack in the decode pool to use for caching, thereby reducing efficiency.
Conversely, with insufficient prefetching, filter threads block while waiting for their source frames to arrive.
We control the size of the prefetch window by capping the number of ActiveGens, which puts backpressure on the decoders, causing them to stall.
Conceptually, this flips decoders between eager and lazy behaviors to balance efficiency and throughput.
Configuring this limit is discussed in our evaluation.

\subsubsection{Implementation Notes}

In prior sections, we propose a multi-threaded rendering engine.
For conciseness, we omit a discussion of the exact synchronization primitives.
Full details are readily available in our source code.
The implementation of any concurrent system poses inherent difficulties in ensuring correctness; we have checked the core concurrency model with $\text{TLA}^{+}$~\cite{lamport1999specifying}, and we include our $\text{TLA}^{+}$ specification in the accompanying source code.

\section{Video Results on Demand} \label{sec:vrod}

Our declarative video transformation rendering engine enables significantly faster rendering by augmenting the control flow to apply optimizations and parallelize rendering.
As discussed in the results, this yields a $2-3\times$ speedup across diverse annotation workloads.
However, this approach is still constrained by the limitations of its components: decoding, filtering, and encoding video frames.
For a truly interactive data exploration experience, rendering optimization alone would be insufficient to deliver the sub-second response times that users perceive as interactive~\cite{liu2014effects}.

Therefore, to serve such an experience, we design Vidformer around partial materialization, avoiding the need to render the entire video.
While \textit{viewing} the resulting video is the end goal, the slow process of writing the entire video to disk is traditionally the usual path of least resistance for most implementations.
Instead of this approach, we render output frames just before they are played on screen, an approach that is compatible with existing \textit{video players} by exposing rendered videos as a \textit{Video on Demand} (VOD) stream.
Vidformer with VOD is a pragmatic approach: VOD standards have broad support in media players due to their use in on-demand streaming services, and VOD protocols are efficient over networks, which is critical for delivering video results to clients on edge devices.
We show how VOD enables practically instantaneous playback of video results.

We use HTTP Live Streaming (HLS)~\cite{rfc8216} in our implementation, but DASH~\cite{ISO23009} is functionally equivalent for this use case.
Frames are partitioned into 2-second segments, and we immediately create a stream manifest file that references them.
When the client requests a segment, Vidformer finds the required frames and passes their declarative representations to our rendering engine, which renders the segment and returns it.
The bottom of~\Cref{fig:vf-methods} shows this in action.
This makes rendering a constant-time operation; only a short segment needs to be rendered to begin playback.
Users can see the entire video in their player and jump to arbitrary points as if the full video were rendered.

\subsection{Incremental Video Results via Live-Streaming}

During our preliminary work integrating Video Results on Demand, we found that rendering was so fast that execution of the visualization script became the bottleneck.
We execute scripts without multimedia operations, running them multiple orders of magnitude faster than in real-time.
However, long videos could cause scripts to take a few seconds, a consequence of Python's relatively slow interpreter and inefficient visualization scripts (e.g., the script in \Cref{fig:cv2-example} filters the dataframe \texttt{df} once per output frame).
This overhead is usually amortized across the entire rendering pipeline, making it negligible, but it became the primary contributor to time-to-playback latency.

To consistently reach sub-second time-to-playback, we incrementally display videos as the visualization script executes.
In our Python/OpenCV implementation, frames are appended to the specification as they are written to the \texttt{cv2.VideoWriter}.
Since the length of the output video is unknown, we use VOD live streaming to convey this to the video player.
Our VOD server marks the manifest as unfinished and progressively adds segments as more frames become available.
Once the \texttt{cv2.VideoWriter} is closed, the manifest reports that the stream is terminated, causing the player to switch to regular on-demand playback.
This is neither a VOD stream, where the whole video is available, nor a live stream, which grows but has a fixed lookback window, but rather a hybrid of the two.
The video stream has a fixed start point, can be appended to, and all segments remain available; HLS refers to this as an ``event stream''.
This incremental live-streaming approach makes playback a constant-time operation, even with slow intermediate queries and visualization scripts.

\subsection{Source Video I/O}

Previously, we've shown code that opens locally stored videos directly using their paths; however, this approach does not work in distributed deployments.
Instead, we access videos in situ, from their underlying storage service, usually within an object store.
In our implementation, we pass all video I/O through Apache OpenDAL~\cite{OpenDAL_Contributors_OpenDAL} to support many different storage protocols.
When integrating with a video database, scripts can use simple paths (e.g., ``\texttt{<video primary key UUID>.mp4}''), which Vidformer maps to the underlying video file locations.
Accessing video files over the network introduces additional latency considerations, even in fast datacenter networks.
Many I/O accesses are repeated as subsequent VOD segments are fetched; therefore, we apply an additional shared I/O block caching layer to eliminate the latency overhead of repeatedly opening and parsing video files.

\subsection{LLM-based Video Querying}\label{sec:llm-querying}

\begin{figure}
    \centering
    \vspace{2mm}
    \includegraphics[width=\linewidth]{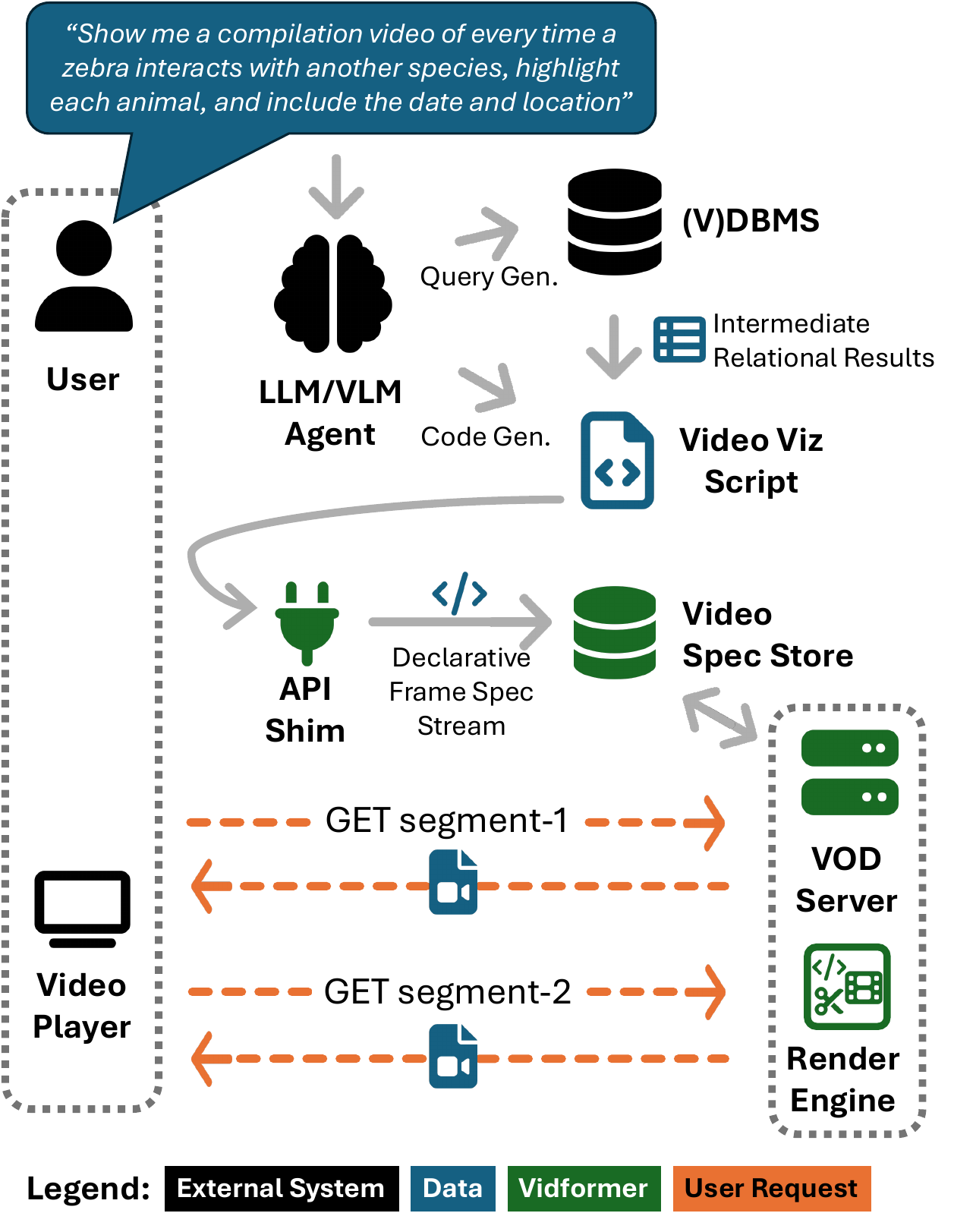}
    \vspace{-5mm}
    \caption{LLM-based Video Querying: When a user issues a ``\textit{show me ...}'' query, an LLM queries a database and writes a video data visualization script. Vidformer runs this script, enabling end-to-end video responses in seconds.}
    \label{fig:llm-querying}
\end{figure}

We detail how Vidformer enables interactive LLM-based video querying.
This serves as an exciting new use case enabled by Vidformer, and an opportunity to detail a productionalized deployment.
In popular culture, the ability to issue commands and view visual media results is a mainstay characteristic of futuristic technology.
Numerous shows and movies, such as Star Trek and Ready Player One, prominently feature characters issuing vocal commands and instantly receiving video or holographic responses.
Modern LLMs can make this a reality by formalizing natural language commands into code, yet this remains impractical given the high latency to render the resulting video.
Vidformer complements LLMs to enable interactive-speed video querying using the approach shown in \Cref{fig:llm-querying}.

We deploy Vidformer as a web service with two endpoints: one for its management API and the other to serve VOD requests.
When a user invokes a ``\textit{show me ...}'' command, the LLM agent generates database queries, a Python/cv2 visualization script, and registers a new specification with the Vidformer service.
This Vidformer service instantly generates a VOD namespace and returns a link to the VOD stream manifest to the user.
The browser then displays a video player, and the VOD client repeatedly polls the manifest file for segments.
Simultaneously, the LLM agent invokes a lightweight virtual machine to execute the video data visualization script in a secure environment.
This script is first augmented to apply the cv2 API layer by replacing \texttt{import cv2} with \texttt{import vidformer.cv2 as cv2}.
Any database query results are copied into the virtual machine, and a frame push API endpoint is provided to the VM, which the Vidformer library iteratively sends frame expressions to as they are written to the \texttt{cv2.VideoWriter} object.
The Vidformer service receives these push requests from the VM and then inserts them into the video spec store.
The push endpoint also applies type checking and security policy.
Type checking ensures that each filter function accepts the provided arguments and that the output resolution and pixel formats of frames are correct.
This ensures that video specifications in the database render correctly and that visualization scripts can quickly identify errors.
Additionally, this endpoint prevents adverse specifications by statically checking that frames comply with security policies, such as restricting intermediate frame resolutions, inlined data value size, and specification tree depth.
Each time the VOD client polls the manifest, the Vidformer service counts the successfully written frames to determine the number of segments to include in the generated manifest.
For example, in a 30 FPS video with 2-second segments, the manifest lists the first segment after the script has written its 60th frame.
The VOD client immediately requests this segment, which is rendered on demand.
A load balancer on the VOD endpoint provides parallel segment rendering, distributing requests across a cluster of rendering host servers or containers.
The load balancer also caches rendered segments, as VOD clients often purge and re-request the same segments, or to allow multiple clients to consume the same stream.
Once the visualization script is complete and the \texttt{cv2.VideoWriter} is closed, the Vidformer API shim makes a final push call to mark the video spec as terminated, and the server then adds an end-of-stream marker to the manifest.
This causes the video player to display the video's ending time and cease polling the manifest.
The VOD player may continue fetching segments until the session ends, at which point a cleanup routine removes the video specification from the specification store.

\subsection{Limitations}

\subsubsection{Scope} The approach in this paper applies only to video transformation for visualization, which produces a video without observing or branching based on pixel values within the visualization script.
Annotation data must be computed externally.
Vidformer is not a general-purpose video processing accelerator; we rely on external systems or the user for this functionality.

\subsubsection{Segment Encoding} An inherent inefficiency of our approach is the added encode/decode overhead in VOD segment encoding.
Passing uncompressed frames from the rendering engine directly into a process's framebuffer would eliminate this overhead; however, uncompressed frames are far too large to transfer over a network, so this is not feasible unless rendering is done directly on the client.
In practice, clients are typically web browsers, making this approach feasible; however, it is an inferior approach at present.
Edge devices tend to be significantly less powerful than servers, and slower networks can make I/O to external video sources difficult.
Most importantly, quick and efficient playback of VOD streams is a major priority for edge device operating systems and web browsers~\cite{10.1145/3212804}.
Deviating from this highly optimized pipeline sacrifices energy efficiency and maximum achievable resolution unless a significant development effort is expended.

\section{Evaluation}

We implemented the Vidformer server and rendering engine in Rust, utilizing FFmpeg's libav* library for low-level video management and OpenCV for most filters.
We implemented handwritten baselines in Python with the cv2 library for OpenCV.
\textbf{Baselines and Vidformer both use FFmpeg's libav* for decoding/encoding and OpenCV for filters, making this a fair comparison.}

\subsubsection*{Datasets}

We use Blender Open Movie's \textit{Tears of Steel}~(ToS), a commonly used reference video for tasks requiring a single video.
It is $1280\times720$, 24 FPS, and $734$ seconds long, and is 56 MB encoded on disk.
We use a low-resolution video for most of our evaluation, as higher resolutions may conceal rendering engine scheduler overhead.
We use downscaled encodings of the 4K original in~\Cref{sec:gpus}.
For tasks requiring a video collection, we use a compilation of 100 documentaries from PBS's \textit{NOVA} \& \textit{FRONTLINE} series, publicly viewable on YouTube.
We selected sources that are $1920\times1080$, 29.97 FPS, and each around 55 minutes long.
These are 70 GB in total.
All videos are H.264-encoded with \texttt{yuv420p} pixel format.

\subsubsection*{Environment}
We used a 2$\times$12-core (48 HT vCPUs) Intel Xeon Gold 6126 system at 2.6 GHz with 384GB DDR4 RAM, running Debian 12.
To accurately measure the impact of threads, we avoid saturating the CPU; since the underlying FFmpeg libav* libraries and their codecs create threads internally, we restrict all of our experiments to 16 total decode \& filter threads.
We used FFmpeg 7.0, OpenCV 4.9, Rust 1.77, and Python 3.11.
We used Postgres for our specification store database and Valkey, an open-source Redis fork, for the source video I/O cache.

\subsection{Results}

\begin{table}[t]
    \centering
    \begin{tabular}{lccc}
        \toprule
        \textbf{Annotations} & \textbf{Baseline} & \textbf{VF} & \textbf{VF+VOD} \\
        \midrule
        Label & 116.09 & 41.32 & 0.28 \\
        Box+Label & 123.20 & 44.21 & 0.25 \\
        BoxCorner+Label & 120.96 & 63.38 & 0.24 \\
        Color+Label & 141.15 & 47.39 & 0.30 \\
        Mask+Label & 144.00 & 51.34 & 0.50 \\
        \bottomrule
    \end{tabular}
    \vspace{2mm}
    \caption{Time-to-playback latency in seconds. Comparing Python \& OpenCV baselines, Vidformer full render, and Vidformer+VOD.}
    \label{tab:sv-anots}
\end{table}

We begin by evaluating simple computer vision annotation tasks.
We use Supervision's~\cite{Roboflow_Supervision} annotators, which internally use OpenCV.
Our ``Label'' task corresponds to \texttt{supervision.LabelAnnotator}.
We annotate the ToS video with precomputed detections from YOLOv8~\cite{7780460} inference with YOLOv8-Seg for MaskAnnotator.
We used eight decode/filter threads with a decode pool size of 100.
Results are shown in \Cref{tab:sv-anots}.
Baseline and VF, for Vidformer, are the time to run the entire visualization script and fully render the resulting video.
With VF+VOD, this is the time to begin the video playback, which only renders one two-second segment.
These results show that the Vidformer rendering engine provides a $2\times$ to $3\times$ speedup of existing workflows.
Viewing the video with VOD begins playback within half a second, resulting in a ${\sim}400\text{--}500\times$ speedup.

\subsubsection{Rendering Parallelism}

\begin{figure}
    \centering
    \includegraphics[width=\linewidth]{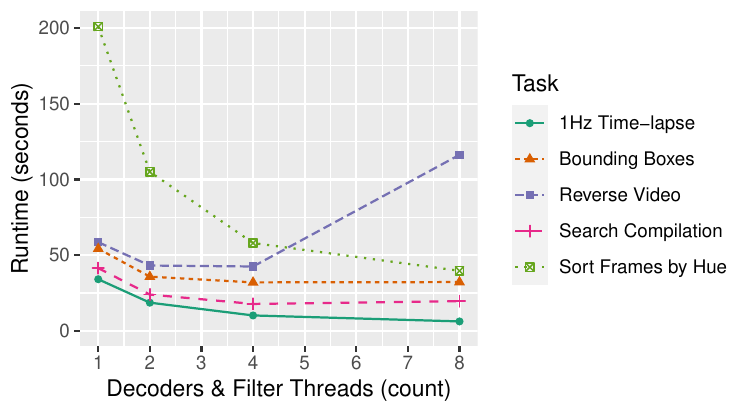}
    \vspace{-6mm}
    \caption{Thread count vs. rendering runtime.}
    \label{fig:core-threads}
\end{figure}

To evaluate how well the Vidformer rendering engine uses additional threads to lower latency, we assess various tasks to stress the rendering engine in diverse ways.
Each of our tasks is on ToS and should be apparent from the name, except ``Search Compilation'', which searches the PBS dataset for a spoken word in the subtitles (we use the word ``river''), creates a compilation, and labels the occurrence number, source video, and timestamp of the source segment.
This results in a 1-minute \& 42-second result, finding 32 segments from 14 different source videos.
We use a decode pool size of 100 frames and a prefetch window of 80.
With this configuration, the decode pool's memory usage is at a maximum of 277MB at 720p and 622MB at 1080p.
The results in \Cref{fig:core-threads} show that Vidformer efficiently distributes workloads across threads, thereby reducing the runtime.
There is one exception to this trend: the ``Reverse Video'' was substantially worse at eight threads than at lower thread counts.
We have verified that this trend also continues to worsen at higher thread counts for this task.
We believe this is due to emergent thrashing behavior in the decoder thread pool.
Frames tend to be decoded in an order close to their temporal position.
Therefore, when frames are needed at the end of a GOP, earlier frames are discarded, causing additional decoders to be spun up.
This issue occurs when the decode order resembles a reverse access order, and it is worst when reversing videos.
This can be alleviated with a decode frame pool large enough to cache one GOP.

\subsubsection{Frame Access Patterns}\label{sec:dense-ap}

\begin{figure}
    \centering
    \includegraphics[width=1.0\linewidth]{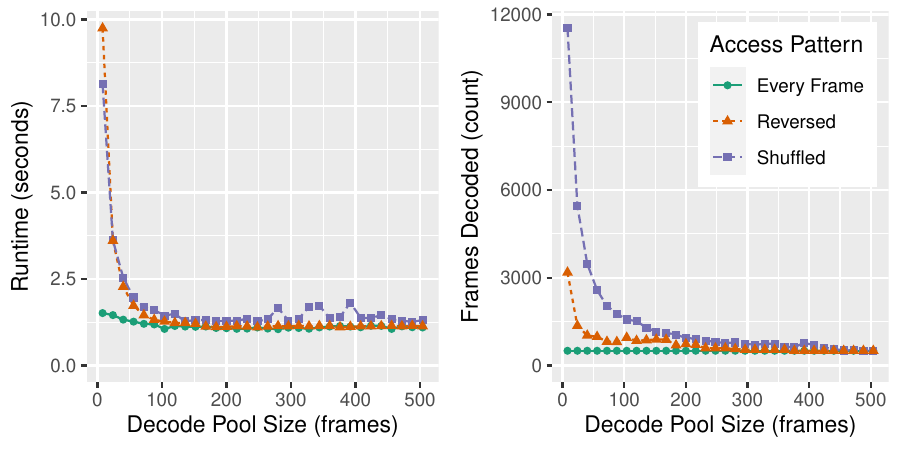}
    \vspace{-3mm}
    \caption{Runtime (left) and the number of frames decoded (right) vs decode pool size for dense frame access patterns. Time to decode 500 frames shown.}
    \label{fig:ap-time-and-decodes}
\end{figure}

We evaluated different frame access patterns to assess our scheduling and frame caching approach.
We evaluate dense access patterns using every frame, reversing a video, and randomly shuffling frames to capture the high, moderate, and low locality access patterns.
We measured the time taken and the number of frames decoded for different decode pool sizes over a 500-frame segment of ToS.
Results are shown in \Cref{fig:ap-time-and-decodes}.
As expected, increasing the decode pool size results in a rapid decline in runtime and the number of frames decoded for each of our tasks.

\begin{figure}
    \centering
    \includegraphics[width=0.90\linewidth]{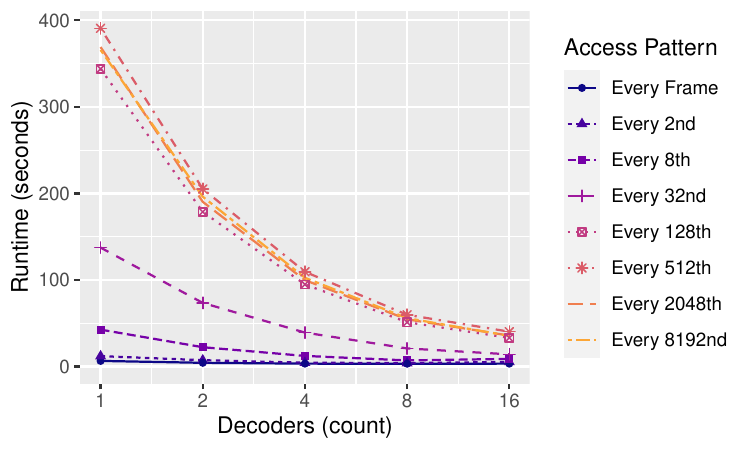}
    \vspace{-2mm}
    \caption{Decoder threads vs. runtime for sparse frame access patterns. Time to extract 1000 frames shown.}
    \label{fig:sparse-ap}
\end{figure}

Accessing frames sparsely is difficult as it suffers from unavoidable frame decode amplification.
At worst, a decoder must decode an entire GOP to retrieve one frame.
In such cases, the only option to speed up transformation is to use additional decoder threads.
We evaluated this behavior over uniform frame access intervals, with strides ranging from 1 to 8192, for a total of 1000 frames.
Since no single video was sufficiently long, we constructed a virtual source video by splicing together every video in our PBS dataset.
This created a virtual video with 9.7 million frames. % 9,714,788 frames
Results, shown in \Cref{fig:sparse-ap}, fall into two primary clusters:
Small strides (1-8 frames) are fast regardless of the number of decode threads, while large strides ($\geq 128$) are slower but execute proportionally to the number of threads.
Moving from 1 to 16 decoders had over a $10\times$ speedup.
With small strides, the single-stream processing approach is dominant, and the decode pool is saturated, waiting for frames from the active GOP.
Larger strides rapidly approach the worst-case decoding of one GOP per output frame; however, this higher GOP/frame ratio allows more GOPs to be represented in the NeedSet at any given time, enabling additional decoder threads to be utilized effectively.

\subsubsection{GPU Acceleration}\label{sec:gpus}

\begin{figure}
    \centering
    \includegraphics[width=1.0\linewidth]{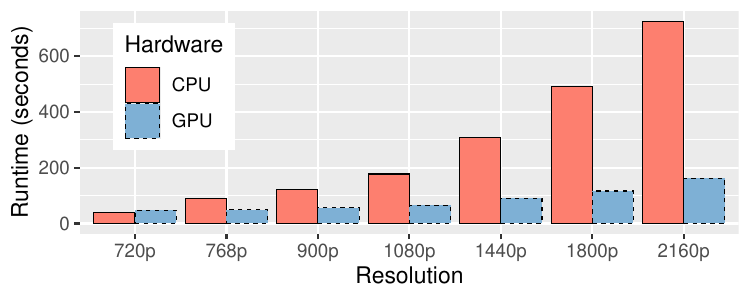}
    \vspace{-7mm}
    \caption{CPU and GPU rendering speeds across resolutions.}
    \label{fig:gpus}
\end{figure}

\Cref{fig:gpus} shows GPU-accelerated Vidformer performance.
We tested the ToS video across common 16:9 resolutions using the prior LabelAnnotator task.
We used an NVIDIA Titan Xp running H.264 NVDEC \& NVENC with the ``\texttt{fast}'' preset.
CPU used the FFmpeg default codec, libx264, with the ``\texttt{ultrafast}'' preset.
The number of filter threads was increased to 24 to alleviate a bottleneck caused by the CPU-based filters.
We ran filters on the CPU in all our experiments, as OpenCV drawing methods are only supported on CPUs.
Full end-to-end GPU acceleration is possible by substituting GPU-accelerated equivalents, such as CV-CUDA~\cite{nvidia_cvcuda}.
For fairness, we do not include this in our evaluation, as it does not provide a pixel-perfect match with our baselines.
The results show GPUs enable consistently fast rendering at higher resolutions, up to $4.49\times$ faster at 4K (2160p).

\subsubsection{LLM-Generated Scripts}

\begin{figure}
    \centering
    \includegraphics[width=1.0\linewidth]{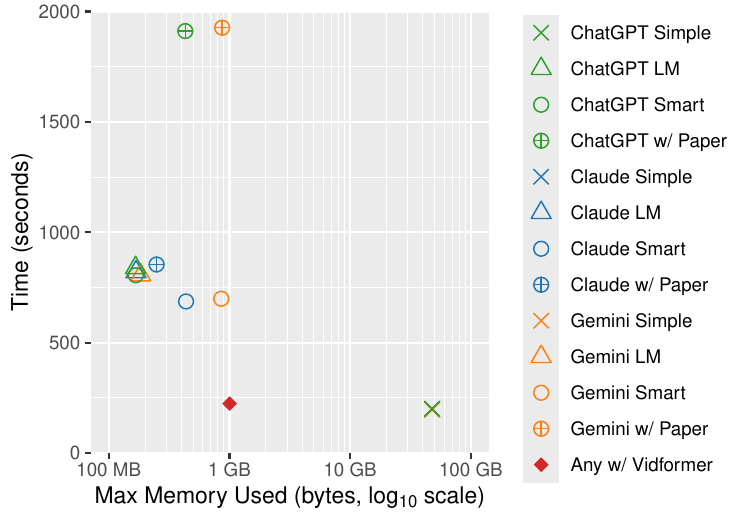}
    \vspace{-6mm}
    \caption{Memory use and runtime of LLM-generated visualization scripts. Different models and prompt guidance are shown, as well as the performance of running \textit{any} of these scripts with Vidformer.}
    \label{fig:llm}
\end{figure}

We asked LLMs for code to sort frames by hue, which is both simple and has clear optimization potential, across OpenAI GPT-5 (ChatGPT), Anthropic Opus 4.1 (Claude), and Google Gemini 2.5 Pro (Gemini).
We asked directly with no guidance (Simple), for a reasonable memory usage (LM), for a highly optimized solution that balances compute and memory (Smart), and a highly optimized solution when provided with the text of \Cref{sec:execution} (w/ Paper).
We included the Vidformer accelerated alternative, which has the same memory and compute profile regardless of which of these scripts we used.
For fairness, we restricted Vidformer to a single thread.
We would like to note that over 700 MB of memory usage is Vidformer server overhead, not actual rendering engine usage.
Our results, in \Cref{fig:llm}, show LLM-generated scripts were wasteful of either time, memory, or both, regardless of the model or prompt.
This demonstrates that LLM-generated video data visualization scripts on larger, longer video datasets would not only be too slow but also often fail due to excessive memory usage.
Applying Vidformer addresses these issues by delivering consistent resource use, allowing LLMs to focus solely on generating correct scripts in the most straightforward way.

\section{Conclusion}

We present \textsc{Vidformer}, a system that enables \textit{instantaneous playback} of video-native query results.
Vidformer models a video data visualization script as a transformation of video data.
We use an API shim to lift imperative scripts into a stream of declarative frame specifications using only a one-line modification.
We design a rendering engine to optimize and parallelize rendering by leveraging the declarative specification.
Additionally, Vidformer serves video results over a VOD protocol, rendering short segments upon request.

Our evaluation demonstrates a $2-3\times$ speedup with our rendering engine and a $400\times$ speedup when combined with VOD streaming.
Furthermore, we find that our rendering engine is effective at utilizing additional threads and memory to reduce latency across a diverse range of tasks.
Even in the worst-case scenario of highly sparse frame accesses, which only benefit from additional threads, we observe linear speedups.
We also evaluated GPU-accelerated rendering and found that GPUs enable higher resolutions with minimal additional latency.

We show how our approach not only accelerates existing video/data visualization pipelines and workflows, but also enables new applications.
Interactive LLM-based natural language querying, where ``\textit{show me ...}'' commands yield instantaneous video results, is not only made possible by Vidformer but also made trivial.
By bringing the speed of a video editor's live preview window to visualization scripts, Vidformer enables truly instant playback of video-native query results.
We release Vidformer as a practical, usable, and open-source system\footnote{\url{https://github.com/ixlab/vidformer}}.

\begin{acks}
This material is based upon work supported by the National Science Foundation under Grant No. 1910356 and the NSF OAC 2118240 Imageomics Institute award.
Icons in \Cref{fig:vf-methods,fig:llm-querying} by Font Awesome (CC BY 4.0).
\end{acks}

\bibliographystyle{ACM-Reference-Format}
\bibliography{main}


\begin{thebibliography}{53}

%%% the Web address.

\ifx \showCODEN    \undefined \def \showCODEN     #1{\unskip}     \fi
\ifx \showDOI      \undefined \def \showDOI       #1{#1}\fi
\ifx \showISBNx    \undefined \def \showISBNx     #1{\unskip}     \fi
\ifx \showISBNxiii \undefined \def \showISBNxiii  #1{\unskip}     \fi
\ifx \showISSN     \undefined \def \showISSN      #1{\unskip}     \fi
\ifx \showLCCN     \undefined \def \showLCCN      #1{\unskip}     \fi
\ifx \shownote     \undefined \def \shownote      #1{#1}          \fi
\ifx \showarticletitle \undefined \def \showarticletitle #1{#1}   \fi
\ifx \showURL      \undefined \def \showURL       {\relax}        \fi
% invisible to TeX
\providecommand\bibfield[2]{#2}
\providecommand\bibinfo[2]{#2}
\providecommand\natexlab[1]{#1}
\providecommand\showeprint[2][]{arXiv:#2}

\bibitem[\protect\citeauthoryear{??}{ape}{2024}]%
        {aperturedb}
 \bibinfo{year}{2024}\natexlab{}.
\newblock \bibinfo{booktitle}{\emph{ApertureDB}}.
\newblock
\urldef\tempurl%
\url{https://github.com/aperture-data/aperturedb-python}
\showURL{%
\tempurl}


\bibitem[\protect\citeauthoryear{??}{noa}{2024a}]%
        {noauthor_creatomate_nodate}
 \bibinfo{year}{2024}\natexlab{a}.
\newblock \bibinfo{title}{{Creatomate}}.
\newblock
\newblock
\urldef\tempurl%
\url{https://creatomate.com/}
\showURL{%
\tempurl}


\bibitem[\protect\citeauthoryear{??}{noa}{2024b}]%
        {noauthor_redotvideorevideo_2024}
 \bibinfo{year}{2024}\natexlab{b}.
\newblock \bibinfo{title}{redotvideo/revideo}.
\newblock
\newblock
\urldef\tempurl%
\url{https://github.com/redotvideo/revideo}
\showURL{%
\tempurl}


\bibitem[\protect\citeauthoryear{??}{noa}{2024c}]%
        {noauthor_remotion_nodate}
 \bibinfo{year}{2024}\natexlab{c}.
\newblock \bibinfo{title}{Remotion}.
\newblock
\newblock
\urldef\tempurl%
\url{https://www.remotion.dev/}
\showURL{%
\tempurl}


\bibitem[\protect\citeauthoryear{??}{noa}{2024d}]%
        {noauthor_shotstack_nodate}
 \bibinfo{year}{2024}\natexlab{d}.
\newblock \bibinfo{title}{Shotstack}.
\newblock
\newblock
\urldef\tempurl%
\url{https://shotstack.io/}
\showURL{%
\tempurl}


\bibitem[\protect\citeauthoryear{??}{noa}{2024e}]%
        {noauthor_videodb_nodate}
 \bibinfo{year}{2024}\natexlab{e}.
\newblock \bibinfo{title}{{VideoDB}}.
\newblock
\newblock
\urldef\tempurl%
\url{https://videodb.io/}
\showURL{%
\tempurl}


\bibitem[\protect\citeauthoryear{Abadi, Agarwal, Barham, Brevdo, Chen, Citro, Corrado, Davis, Dean, Devin, Ghemawat, Goodfellow, Harp, Irving, Isard, Jia, Jozefowicz, Kaiser, Kudlur, Levenberg, Man\'{e}, Monga, Moore, Murray, Olah, Schuster, Shlens, Steiner, Sutskever, Talwar, Tucker, Vanhoucke, Vasudevan, Vi\'{e}gas, Vinyals, Warden, Wattenberg, Wicke, Yu, and Zheng}{Abadi et~al\mbox{.}}{2015}]%
        {tensorflow2015-whitepaper}
\bibfield{author}{\bibinfo{person}{Mart\'{i}n Abadi}, \bibinfo{person}{Ashish Agarwal}, \bibinfo{person}{Paul Barham}, \bibinfo{person}{Eugene Brevdo}, \bibinfo{person}{Zhifeng Chen}, \bibinfo{person}{Craig Citro}, \bibinfo{person}{Greg~S. Corrado}, \bibinfo{person}{Andy Davis}, \bibinfo{person}{Jeffrey Dean}, \bibinfo{person}{Matthieu Devin}, \bibinfo{person}{Sanjay Ghemawat}, \bibinfo{person}{Ian Goodfellow}, \bibinfo{person}{Andrew Harp}, \bibinfo{person}{Geoffrey Irving}, \bibinfo{person}{Michael Isard}, \bibinfo{person}{Yangqing Jia}, \bibinfo{person}{Rafal Jozefowicz}, \bibinfo{person}{Lukasz Kaiser}, \bibinfo{person}{Manjunath Kudlur}, \bibinfo{person}{Josh Levenberg}, \bibinfo{person}{Dandelion Man\'{e}}, \bibinfo{person}{Rajat Monga}, \bibinfo{person}{Sherry Moore}, \bibinfo{person}{Derek Murray}, \bibinfo{person}{Chris Olah}, \bibinfo{person}{Mike Schuster}, \bibinfo{person}{Jonathon Shlens}, \bibinfo{person}{Benoit Steiner}, \bibinfo{person}{Ilya Sutskever}, \bibinfo{person}{Kunal Talwar},
  \bibinfo{person}{Paul Tucker}, \bibinfo{person}{Vincent Vanhoucke}, \bibinfo{person}{Vijay Vasudevan}, \bibinfo{person}{Fernanda Vi\'{e}gas}, \bibinfo{person}{Oriol Vinyals}, \bibinfo{person}{Pete Warden}, \bibinfo{person}{Martin Wattenberg}, \bibinfo{person}{Martin Wicke}, \bibinfo{person}{Yuan Yu}, {and} \bibinfo{person}{Xiaoqiang Zheng}.} \bibinfo{year}{2015}\natexlab{}.
\newblock \bibinfo{title}{{TensorFlow}: Large-Scale Machine Learning on Heterogeneous Systems}.
\newblock
\newblock
\urldef\tempurl%
\url{https://www.tensorflow.org/}
\showURL{%
\tempurl}
\newblock
\shownote{Software available from tensorflow.org.}


\bibitem[\protect\citeauthoryear{Abdallah, Griwodz, Chen, Simon, Wang, and Hsu}{Abdallah et~al\mbox{.}}{2018}]%
        {10.1145/3212804}
\bibfield{author}{\bibinfo{person}{Maha Abdallah}, \bibinfo{person}{Carsten Griwodz}, \bibinfo{person}{Kuan-Ta Chen}, \bibinfo{person}{Gwendal Simon}, \bibinfo{person}{Pin-Chun Wang}, {and} \bibinfo{person}{Cheng-Hsin Hsu}.} \bibinfo{year}{2018}\natexlab{}.
\newblock \showarticletitle{Delay-Sensitive Video Computing in the Cloud: A Survey}.
\newblock \bibinfo{journal}{\emph{ACM Trans. Multimedia Comput. Commun. Appl.}} \bibinfo{volume}{14}, \bibinfo{number}{3s}, Article \bibinfo{articleno}{54} (\bibinfo{date}{June} \bibinfo{year}{2018}), \bibinfo{numpages}{29}~pages.
\newblock
\showISSN{1551-6857}
\urldef\tempurl%
\url{https://doi.org/10.1145/3212804}
\showDOI{\tempurl}


\bibitem[\protect\citeauthoryear{Andersen, Chang, and Felleisen}{Andersen et~al\mbox{.}}{2017}]%
        {andersen_super_2017}
\bibfield{author}{\bibinfo{person}{Leif Andersen}, \bibinfo{person}{Stephen Chang}, {and} \bibinfo{person}{Matthias Felleisen}.} \bibinfo{year}{2017}\natexlab{}.
\newblock \showarticletitle{Super 8 languages for making movies (functional pearl)}.
\newblock \bibinfo{journal}{\emph{Proceedings of the ACM on Programming Languages}} \bibinfo{volume}{1}, \bibinfo{number}{ICFP} (\bibinfo{date}{aug} \bibinfo{year}{2017}), \bibinfo{pages}{30:1--30:29}.
\newblock
\urldef\tempurl%
\url{https://doi.org/10.1145/3110274}
\showDOI{\tempurl}


\bibitem[\protect\citeauthoryear{Arslan}{Arslan}{2002}]%
        {arslan_semantic_2002}
\bibfield{author}{\bibinfo{person}{Umut Arslan}.} \bibinfo{year}{2002}\natexlab{}.
\newblock \emph{\bibinfo{title}{A {Semantic} {Data} {Model} and {Query} {Language} for {Video} {Databases}}}.
\newblock {PhD} {Thesis}.
\newblock
\newblock
\shownote{ISBN: 9798426865716 Publication Title: PQDT - Global.}


\bibitem[\protect\citeauthoryear{Bocconi}{Bocconi}{2004}]%
        {bocconi_semantic-aware_2004}
\bibfield{author}{\bibinfo{person}{Stefano Bocconi}.} \bibinfo{year}{2004}\natexlab{}.
\newblock \showarticletitle{Semantic-aware automatic video editing}. In \bibinfo{booktitle}{\emph{Proceedings of the 12th annual {ACM} international conference on {Multimedia}}}. \bibinfo{publisher}{ACM}, \bibinfo{address}{New York NY USA}, \bibinfo{pages}{971--972}.
\newblock
\showISBNx{978-1-58113-893-1}
\urldef\tempurl%
\url{https://doi.org/10.1145/1027527.1027753}
\showDOI{\tempurl}


\bibitem[\protect\citeauthoryear{Bradski}{Bradski}{2000}]%
        {opencv_library}
\bibfield{author}{\bibinfo{person}{G. Bradski}.} \bibinfo{year}{2000}\natexlab{}.
\newblock \showarticletitle{{The OpenCV Library}}.
\newblock \bibinfo{journal}{\emph{Dr. Dobb's Journal of Software Tools}} (\bibinfo{year}{2000}).
\newblock


\bibitem[\protect\citeauthoryear{Catarci, Donderler, Saykol, Ulusoy, and Gudukbay}{Catarci et~al\mbox{.}}{2003}]%
        {catarci_bilvideo_2003}
\bibfield{author}{\bibinfo{person}{T Catarci}, \bibinfo{person}{ME Donderler}, \bibinfo{person}{E Saykol}, \bibinfo{person}{O Ulusoy}, {and} \bibinfo{person}{U Gudukbay}.} \bibinfo{year}{2003}\natexlab{}.
\newblock \showarticletitle{{BilVideo}: {A} video database management system}.
\newblock \bibinfo{journal}{\emph{IEEE MultiMedia}} \bibinfo{volume}{10}, \bibinfo{number}{1} (\bibinfo{year}{2003}), \bibinfo{pages}{66--70}.
\newblock
\urldef\tempurl%
\url{https://doi.org/10.1109/MMUL.2003.1167924}
\showDOI{\tempurl}
\newblock
\shownote{Publisher: IEEE.}


\bibitem[\protect\citeauthoryear{Contributors}{Contributors}{[n.d.]}]%
        {OpenDAL_Contributors_OpenDAL}
\bibfield{author}{\bibinfo{person}{OpenDAL Contributors}.} \bibinfo{year}{[n.d.]}\natexlab{}.
\newblock \bibinfo{booktitle}{\emph{{OpenDAL}}}.
\newblock
\urldef\tempurl%
\url{https://github.com/apache/opendal}
\showURL{%
\tempurl}


\bibitem[\protect\citeauthoryear{Corporation}{Corporation}{2024}]%
        {nvidia_cvcuda}
\bibfield{author}{\bibinfo{person}{NVIDIA Corporation}.} \bibinfo{year}{2024}\natexlab{}.
\newblock \bibinfo{title}{CV-CUDA}.
\newblock \bibinfo{howpublished}{\url{https://developer.nvidia.com/cv-cuda}}.
\newblock
\newblock
\shownote{Accessed: 2024-11-18.}


\bibitem[\protect\citeauthoryear{{FFmpeg Developers}}{{FFmpeg Developers}}{2024}]%
        {ffmpeg_developers_ffmpeg_2024}
\bibfield{author}{\bibinfo{person}{{FFmpeg Developers}}.} \bibinfo{year}{2024}\natexlab{}.
\newblock \bibinfo{title}{ffmpeg tool}.
\newblock
\newblock
\urldef\tempurl%
\url{http://ffmpeg.org/}
\showURL{%
\tempurl}


\bibitem[\protect\citeauthoryear{Finstad}{Finstad}{2024}]%
        {finstad_mifieditly_2024}
\bibfield{author}{\bibinfo{person}{Mikael Finstad}.} \bibinfo{year}{2024}\natexlab{}.
\newblock \bibinfo{title}{mifi/editly}.
\newblock
\newblock
\urldef\tempurl%
\url{https://github.com/mifi/editly}
\showURL{%
\tempurl}
\newblock
\shownote{original-date: 2020-04-15.}


\bibitem[\protect\citeauthoryear{Fouse, Weibel, Hutchins, and Hollan}{Fouse et~al\mbox{.}}{2011}]%
        {fouse_chronoviz_2011}
\bibfield{author}{\bibinfo{person}{Adam Fouse}, \bibinfo{person}{Nadir Weibel}, \bibinfo{person}{Edwin Hutchins}, {and} \bibinfo{person}{James~D. Hollan}.} \bibinfo{year}{2011}\natexlab{}.
\newblock \showarticletitle{{ChronoViz}: a system for supporting navigation of time-coded data}. In \bibinfo{booktitle}{\emph{{CHI} '11 {Extended} {Abstracts} on {Human} {Factors} in {Computing} {Systems}}} \emph{(\bibinfo{series}{{CHI} {EA} '11})}. \bibinfo{publisher}{Association for Computing Machinery}, \bibinfo{address}{New York, NY, USA}, \bibinfo{pages}{299--304}.
\newblock
\showISBNx{978-1-4503-0268-5}
\urldef\tempurl%
\url{https://doi.org/10.1145/1979742.1979706}
\showDOI{\tempurl}


\bibitem[\protect\citeauthoryear{Frey, Chi, Yang, and Essa}{Frey et~al\mbox{.}}{2021}]%
        {frey_automatic_2021}
\bibfield{author}{\bibinfo{person}{Nathan Frey}, \bibinfo{person}{Peggy Chi}, \bibinfo{person}{Weilong Yang}, {and} \bibinfo{person}{Irfan Essa}.} \bibinfo{year}{2021}\natexlab{}.
\newblock \bibinfo{title}{Automatic {Non}-{Linear} {Video} {Editing} {Transfer}}.
\newblock
\newblock
\urldef\tempurl%
\url{https://doi.org/10.48550/arXiv.2105.06988}
\showDOI{\tempurl}
\newblock
\shownote{arXiv:2105.06988 [cs].}


\bibitem[\protect\citeauthoryear{Garcia, López, Gortázar, Gallego, and Carella}{Garcia et~al\mbox{.}}{2017}]%
        {garcia_nubomedia_2017}
\bibfield{author}{\bibinfo{person}{Boni Garcia}, \bibinfo{person}{Luis López}, \bibinfo{person}{Francisco Gortázar}, \bibinfo{person}{Micael Gallego}, {and} \bibinfo{person}{Giuseppe~Antonio Carella}.} \bibinfo{year}{2017}\natexlab{}.
\newblock \showarticletitle{{NUBOMEDIA}: {The} {First} {Open} {Source} {WebRTC} {PaaS}}. In \bibinfo{booktitle}{\emph{Proceedings of the 25th {ACM} international conference on {Multimedia}}} \emph{(\bibinfo{series}{{MM} '17})}. \bibinfo{publisher}{Association for Computing Machinery}, \bibinfo{address}{New York, NY, USA}, \bibinfo{pages}{1205--1208}.
\newblock
\showISBNx{978-1-4503-4906-2}
\urldef\tempurl%
\url{https://doi.org/10.1145/3123266.3129392}
\showDOI{\tempurl}


\bibitem[\protect\citeauthoryear{Haas and Hellerstein}{Haas and Hellerstein}{1999}]%
        {haas_ripple_1999}
\bibfield{author}{\bibinfo{person}{Peter~J. Haas} {and} \bibinfo{person}{Joseph~M. Hellerstein}.} \bibinfo{year}{1999}\natexlab{}.
\newblock \showarticletitle{Ripple joins for online aggregation}. In \bibinfo{booktitle}{\emph{Proceedings of the 1999 ACM SIGMOD International Conference on Management of Data}} (Philadelphia, Pennsylvania, USA) \emph{(\bibinfo{series}{SIGMOD '99})}. \bibinfo{publisher}{Association for Computing Machinery}, \bibinfo{address}{New York, NY, USA}, \bibinfo{pages}{287–298}.
\newblock
\showISBNx{1581130848}
\urldef\tempurl%
\url{https://doi.org/10.1145/304182.304208}
\showDOI{\tempurl}


\bibitem[\protect\citeauthoryear{Hellerstein, Haas, and Wang}{Hellerstein et~al\mbox{.}}{1997}]%
        {hellerstein_online_1997}
\bibfield{author}{\bibinfo{person}{Joseph~M. Hellerstein}, \bibinfo{person}{Peter~J. Haas}, {and} \bibinfo{person}{Helen~J. Wang}.} \bibinfo{year}{1997}\natexlab{}.
\newblock \showarticletitle{Online aggregation}.
\newblock \bibinfo{journal}{\emph{SIGMOD Rec.}} \bibinfo{volume}{26}, \bibinfo{number}{2} (\bibinfo{date}{June} \bibinfo{year}{1997}), \bibinfo{pages}{171–182}.
\newblock
\showISSN{0163-5808}
\urldef\tempurl%
\url{https://doi.org/10.1145/253262.253291}
\showDOI{\tempurl}


\bibitem[\protect\citeauthoryear{Hwang and Subrahmanian}{Hwang and Subrahmanian}{1996}]%
        {hwang_querying_1996}
\bibfield{author}{\bibinfo{person}{Eenjun Hwang} {and} \bibinfo{person}{V.~S. Subrahmanian}.} \bibinfo{year}{1996}\natexlab{}.
\newblock \showarticletitle{Querying {Video} {Libraries}*}.
\newblock \bibinfo{journal}{\emph{Journal of Visual Communication and Image Representation}} \bibinfo{volume}{7}, \bibinfo{number}{1} (\bibinfo{date}{mar} \bibinfo{year}{1996}), \bibinfo{pages}{44--60}.
\newblock
\showISSN{1047-3203}
\urldef\tempurl%
\url{https://doi.org/10.1006/jvci.1996.0005}
\showDOI{\tempurl}


\bibitem[\protect\citeauthoryear{ISO/IEC 23009-1:2022}{ISO/IEC 23009-1:2022}{2022}]%
        {ISO23009}
ISO/IEC 23009-1:2022 \bibinfo{year}{2022}\natexlab{}.
\newblock \bibinfo{booktitle}{\emph{{Information technology — Dynamic adaptive streaming over HTTP (DASH)}}}.
\newblock \bibinfo{type}{Standard}. \bibinfo{institution}{International Organization for Standardization}.
\newblock


\bibitem[\protect\citeauthoryear{Jermaine, Arumugam, Pol, and Dobra}{Jermaine et~al\mbox{.}}{2008}]%
        {10.1145/1412331.1412335}
\bibfield{author}{\bibinfo{person}{Chris Jermaine}, \bibinfo{person}{Subramanian Arumugam}, \bibinfo{person}{Abhijit Pol}, {and} \bibinfo{person}{Alin Dobra}.} \bibinfo{year}{2008}\natexlab{}.
\newblock \showarticletitle{Scalable approximate query processing with the DBO engine}.
\newblock \bibinfo{journal}{\emph{ACM Trans. Database Syst.}} \bibinfo{volume}{33}, \bibinfo{number}{4}, Article \bibinfo{articleno}{23} (\bibinfo{date}{Dec.} \bibinfo{year}{2008}), \bibinfo{numpages}{54}~pages.
\newblock
\showISSN{0362-5915}
\urldef\tempurl%
\url{https://doi.org/10.1145/1412331.1412335}
\showDOI{\tempurl}


\bibitem[\protect\citeauthoryear{Kakkar, Cao, Chunduri, Xu, Vyalla, Dintyala, Prabakaran, Bang, Sengupta, Ravichandran, Sivakumar, Rajoria, Raju, Aggarwal, Shah, Garg, Suman, Kalluraya, Mitra, Payani, Lu, Ramachandran, and Arulraj}{Kakkar et~al\mbox{.}}{2023}]%
        {kakkar_eva_2023}
\bibfield{author}{\bibinfo{person}{Gaurav~Tarlok Kakkar}, \bibinfo{person}{Jiashen Cao}, \bibinfo{person}{Pramod Chunduri}, \bibinfo{person}{Zhuangdi Xu}, \bibinfo{person}{Suryatej~Reddy Vyalla}, \bibinfo{person}{Prashanth Dintyala}, \bibinfo{person}{Anirudh Prabakaran}, \bibinfo{person}{Jaeho Bang}, \bibinfo{person}{Aubhro Sengupta}, \bibinfo{person}{Kaushik Ravichandran}, \bibinfo{person}{Ishwarya Sivakumar}, \bibinfo{person}{Aryan Rajoria}, \bibinfo{person}{Ashmita Raju}, \bibinfo{person}{Tushar Aggarwal}, \bibinfo{person}{Abdullah Shah}, \bibinfo{person}{Sanjana Garg}, \bibinfo{person}{Shashank Suman}, \bibinfo{person}{Myna~Prasanna Kalluraya}, \bibinfo{person}{Subrata Mitra}, \bibinfo{person}{Ali Payani}, \bibinfo{person}{Yao Lu}, \bibinfo{person}{Umakishore Ramachandran}, {and} \bibinfo{person}{Joy Arulraj}.} \bibinfo{year}{2023}\natexlab{}.
\newblock \showarticletitle{{EVA}: {An} {End}-to-{End} {Exploratory} {Video} {Analytics} {System}}. In \bibinfo{booktitle}{\emph{Proceedings of the {Seventh} {Workshop} on {Data} {Management} for {End}-to-{End} {Machine} {Learning}}}. \bibinfo{publisher}{ACM}, \bibinfo{address}{Seattle WA USA}, \bibinfo{pages}{1--5}.
\newblock
\showISBNx{9798400702044}
\urldef\tempurl%
\url{https://doi.org/10.1145/3595360.3595858}
\showDOI{\tempurl}


\bibitem[\protect\citeauthoryear{Kang, Bailis, and Zaharia}{Kang et~al\mbox{.}}{2019}]%
        {kang_challenges_2019}
\bibfield{author}{\bibinfo{person}{Daniel Kang}, \bibinfo{person}{Peter Bailis}, {and} \bibinfo{person}{Matei Zaharia}.} \bibinfo{year}{2019}\natexlab{}.
\newblock \showarticletitle{Challenges and {Opportunities} in {DNN}-{Based} {Video} {Analytics}: {A} {Demonstration} of the {BlazeIt} {Video} {Query} {Engine}.}. In \bibinfo{booktitle}{\emph{{CIDR}}}.
\newblock


\bibitem[\protect\citeauthoryear{Kang, Romero, Bailis, Kozyrakis, and Zaharia}{Kang et~al\mbox{.}}{2022}]%
        {kang_viva_2022}
\bibfield{author}{\bibinfo{person}{Daniel Kang}, \bibinfo{person}{Francisco Romero}, \bibinfo{person}{Peter Bailis}, \bibinfo{person}{Christos Kozyrakis}, {and} \bibinfo{person}{Matei Zaharia}.} \bibinfo{year}{2022}\natexlab{}.
\newblock \showarticletitle{{VIVA}: an end-to-end system for interactive video analytics}. \bibinfo{publisher}{CIDR}.
\newblock


\bibitem[\protect\citeauthoryear{Kirillov, Mintun, Ravi, Mao, Rolland, Gustafson, Xiao, Whitehead, Berg, Lo, Dollar, and Girshick}{Kirillov et~al\mbox{.}}{2023}]%
        {Kirillov_2023_ICCV}
\bibfield{author}{\bibinfo{person}{Alexander Kirillov}, \bibinfo{person}{Eric Mintun}, \bibinfo{person}{Nikhila Ravi}, \bibinfo{person}{Hanzi Mao}, \bibinfo{person}{Chloe Rolland}, \bibinfo{person}{Laura Gustafson}, \bibinfo{person}{Tete Xiao}, \bibinfo{person}{Spencer Whitehead}, \bibinfo{person}{Alexander~C. Berg}, \bibinfo{person}{Wan-Yen Lo}, \bibinfo{person}{Piotr Dollar}, {and} \bibinfo{person}{Ross Girshick}.} \bibinfo{year}{2023}\natexlab{}.
\newblock \showarticletitle{Segment Anything}. In \bibinfo{booktitle}{\emph{Proceedings of the IEEE/CVF International Conference on Computer Vision (ICCV)}}. \bibinfo{pages}{4015--4026}.
\newblock


\bibitem[\protect\citeauthoryear{Kittivorawong, Ge, Helal, and Cheung}{Kittivorawong et~al\mbox{.}}{2023}]%
        {kittivorawong_spatialyze_2023}
\bibfield{author}{\bibinfo{person}{Chanwut Kittivorawong}, \bibinfo{person}{Yongming Ge}, \bibinfo{person}{Yousef Helal}, {and} \bibinfo{person}{Alvin Cheung}.} \bibinfo{year}{2023}\natexlab{}.
\newblock \bibinfo{title}{Spatialyze: {A} {Geospatial} {Video} {Analytics} {System} with {Spatial}-{Aware} {Optimizations}}.
\newblock
\newblock
\urldef\tempurl%
\url{http://arxiv.org/abs/2308.03276}
\showURL{%
\tempurl}
\newblock
\shownote{arXiv:2308.03276 [cs].}


\bibitem[\protect\citeauthoryear{Klokmose, Remy, Kristensen, Bagge, Beaudouin-Lafon, and Mackay}{Klokmose et~al\mbox{.}}{2019}]%
        {klokmose_videostrates_2019}
\bibfield{author}{\bibinfo{person}{Clemens~N. Klokmose}, \bibinfo{person}{Christian Remy}, \bibinfo{person}{Janus~Bager Kristensen}, \bibinfo{person}{Rolf Bagge}, \bibinfo{person}{Michel Beaudouin-Lafon}, {and} \bibinfo{person}{Wendy Mackay}.} \bibinfo{year}{2019}\natexlab{}.
\newblock \showarticletitle{Videostrates: {Collaborative}, {Distributed} and {Programmable} {Video} {Manipulation}}. In \bibinfo{booktitle}{\emph{Proceedings of the 32nd {Annual} {ACM} {Symposium} on {User} {Interface} {Software} and {Technology}}} \emph{(\bibinfo{series}{{UIST} '19})}. \bibinfo{publisher}{Association for Computing Machinery}, \bibinfo{address}{New York, NY, USA}, \bibinfo{pages}{233--247}.
\newblock
\showISBNx{978-1-4503-6816-2}
\urldef\tempurl%
\url{https://doi.org/10.1145/3332165.3347912}
\showDOI{\tempurl}


\bibitem[\protect\citeauthoryear{Lamport}{Lamport}{1999}]%
        {lamport1999specifying}
\bibfield{author}{\bibinfo{person}{Leslie Lamport}.} \bibinfo{year}{1999}\natexlab{}.
\newblock \showarticletitle{Specifying Concurrent Systems with TLA+}.
\newblock \bibinfo{journal}{\emph{Calculational System Design}} (\bibinfo{date}{April} \bibinfo{year}{1999}), \bibinfo{pages}{183--247}.
\newblock
\urldef\tempurl%
\url{https://www.microsoft.com/en-us/research/publication/specifying-concurrent-systems-tla/}
\showURL{%
\tempurl}


\bibitem[\protect\citeauthoryear{Liu and Heer}{Liu and Heer}{2014}]%
        {liu2014effects}
\bibfield{author}{\bibinfo{person}{Zhicheng Liu} {and} \bibinfo{person}{Jeffrey Heer}.} \bibinfo{year}{2014}\natexlab{}.
\newblock \showarticletitle{The effects of interactive latency on exploratory visual analysis}.
\newblock \bibinfo{journal}{\emph{IEEE transactions on visualization and computer graphics}} \bibinfo{volume}{20}, \bibinfo{number}{12} (\bibinfo{year}{2014}), \bibinfo{pages}{2122--2131}.
\newblock


\bibitem[\protect\citeauthoryear{Mackay}{Mackay}{1989}]%
        {mackay_eva_1989}
\bibfield{author}{\bibinfo{person}{W.~E. Mackay}.} \bibinfo{year}{1989}\natexlab{}.
\newblock \showarticletitle{{EVA}: an experimental video annotator for symbolic analysis of video data}.
\newblock \bibinfo{journal}{\emph{ACM SIGCHI Bulletin}} \bibinfo{volume}{21}, \bibinfo{number}{2} (\bibinfo{date}{oct} \bibinfo{year}{1989}), \bibinfo{pages}{68--71}.
\newblock
\showISSN{0736-6906}
\urldef\tempurl%
\url{https://doi.org/10.1145/70609.70617}
\showDOI{\tempurl}


\bibitem[\protect\citeauthoryear{Mackay and Beaudouin-Lafon}{Mackay and Beaudouin-Lafon}{1998}]%
        {mackay_diva_1998}
\bibfield{author}{\bibinfo{person}{Wendy~E. Mackay} {and} \bibinfo{person}{Michel Beaudouin-Lafon}.} \bibinfo{year}{1998}\natexlab{}.
\newblock \showarticletitle{{DIVA}: exploratory data analysis with multimedia streams}. In \bibinfo{booktitle}{\emph{Proceedings of the {SIGCHI} {Conference} on {Human} {Factors} in {Computing} {Systems}}} \emph{(\bibinfo{series}{{CHI} '98})}. \bibinfo{publisher}{ACM Press/Addison-Wesley Publishing Co.}, \bibinfo{address}{USA}, \bibinfo{pages}{416--423}.
\newblock
\showISBNx{978-0-201-30987-4}
\urldef\tempurl%
\url{https://doi.org/10.1145/274644.274701}
\showDOI{\tempurl}


\bibitem[\protect\citeauthoryear{Mackay and Davenport}{Mackay and Davenport}{1989}]%
        {mackay_virtual_1989}
\bibfield{author}{\bibinfo{person}{Wendy~E. Mackay} {and} \bibinfo{person}{Glorianna Davenport}.} \bibinfo{year}{1989}\natexlab{}.
\newblock \showarticletitle{Virtual video editing in interactive multimedia applications}.
\newblock \bibinfo{journal}{\emph{Commun. ACM}} \bibinfo{volume}{32}, \bibinfo{number}{7} (\bibinfo{date}{jul} \bibinfo{year}{1989}), \bibinfo{pages}{802--810}.
\newblock
\showISSN{0001-0782}
\urldef\tempurl%
\url{https://doi.org/10.1145/65445.65447}
\showDOI{\tempurl}


\bibitem[\protect\citeauthoryear{Matthews, Gloor, and Makedon}{Matthews et~al\mbox{.}}{1993}]%
        {matthews_videoscheme_1993}
\bibfield{author}{\bibinfo{person}{James Matthews}, \bibinfo{person}{Peter Gloor}, {and} \bibinfo{person}{Fillia Makedon}.} \bibinfo{year}{1993}\natexlab{}.
\newblock \showarticletitle{{VideoScheme}: a programmable video editing systems for automation and media recognition}. In \bibinfo{booktitle}{\emph{Proceedings of the first {ACM} international conference on {Multimedia}}} \emph{(\bibinfo{series}{{MULTIMEDIA} '93})}. \bibinfo{publisher}{Association for Computing Machinery}, \bibinfo{address}{New York, NY, USA}, \bibinfo{pages}{419--426}.
\newblock
\showISBNx{978-0-89791-596-0}
\urldef\tempurl%
\url{https://doi.org/10.1145/166266.168442}
\showDOI{\tempurl}


\bibitem[\protect\citeauthoryear{Niedermayer, Rice, and Martinez}{Niedermayer et~al\mbox{.}}{2021}]%
        {rfc9043}
\bibfield{author}{\bibinfo{person}{Michael Niedermayer}, \bibinfo{person}{Dave Rice}, {and} \bibinfo{person}{Jérôme Martinez}.} \bibinfo{year}{2021}\natexlab{}.
\newblock \bibinfo{title}{{FFV1 Video Coding Format Versions 0, 1, and 3}}.
\newblock \bibinfo{howpublished}{RFC 9043}.
\newblock
\urldef\tempurl%
\url{https://doi.org/10.17487/RFC9043}
\showDOI{\tempurl}


\bibitem[\protect\citeauthoryear{Okuta, Unno, Nishino, Hido, and Loomis}{Okuta et~al\mbox{.}}{2017}]%
        {cupy_learningsys2017}
\bibfield{author}{\bibinfo{person}{Ryosuke Okuta}, \bibinfo{person}{Yuya Unno}, \bibinfo{person}{Daisuke Nishino}, \bibinfo{person}{Shohei Hido}, {and} \bibinfo{person}{Crissman Loomis}.} \bibinfo{year}{2017}\natexlab{}.
\newblock \showarticletitle{CuPy: A NumPy-Compatible Library for NVIDIA GPU Calculations}. In \bibinfo{booktitle}{\emph{Proceedings of Workshop on Machine Learning Systems (LearningSys) in The Thirty-first Annual Conference on Neural Information Processing Systems (NIPS)}}.
\newblock
\urldef\tempurl%
\url{http://learningsys.org/nips17/assets/papers/paper_16.pdf}
\showURL{%
\tempurl}


\bibitem[\protect\citeauthoryear{Pantos and May}{Pantos and May}{2017}]%
        {rfc8216}
\bibfield{author}{\bibinfo{person}{Roger Pantos} {and} \bibinfo{person}{William May}.} \bibinfo{year}{2017}\natexlab{}.
\newblock \bibinfo{title}{{HTTP Live Streaming}}.
\newblock \bibinfo{howpublished}{RFC 8216}.
\newblock
\urldef\tempurl%
\url{https://doi.org/10.17487/RFC8216}
\showDOI{\tempurl}


\bibitem[\protect\citeauthoryear{Paszke, Gross, Massa, Lerer, Bradbury, Chanan, Killeen, Lin, Gimelshein, Antiga, Desmaison, Köpf, Yang, DeVito, Raison, Tejani, Chilamkurthy, Steiner, Fang, Bai, and Chintala}{Paszke et~al\mbox{.}}{2019}]%
        {paszke2019pytorchimperativestylehighperformance}
\bibfield{author}{\bibinfo{person}{Adam Paszke}, \bibinfo{person}{Sam Gross}, \bibinfo{person}{Francisco Massa}, \bibinfo{person}{Adam Lerer}, \bibinfo{person}{James Bradbury}, \bibinfo{person}{Gregory Chanan}, \bibinfo{person}{Trevor Killeen}, \bibinfo{person}{Zeming Lin}, \bibinfo{person}{Natalia Gimelshein}, \bibinfo{person}{Luca Antiga}, \bibinfo{person}{Alban Desmaison}, \bibinfo{person}{Andreas Köpf}, \bibinfo{person}{Edward Yang}, \bibinfo{person}{Zach DeVito}, \bibinfo{person}{Martin Raison}, \bibinfo{person}{Alykhan Tejani}, \bibinfo{person}{Sasank Chilamkurthy}, \bibinfo{person}{Benoit Steiner}, \bibinfo{person}{Lu Fang}, \bibinfo{person}{Junjie Bai}, {and} \bibinfo{person}{Soumith Chintala}.} \bibinfo{year}{2019}\natexlab{}.
\newblock \bibinfo{title}{PyTorch: An Imperative Style, High-Performance Deep Learning Library}.
\newblock
\newblock
\showeprint[arxiv]{1912.01703}~[cs.LG]
\urldef\tempurl%
\url{https://arxiv.org/abs/1912.01703}
\showURL{%
\tempurl}


\bibitem[\protect\citeauthoryear{Petersohn, Macke, Xin, Ma, Lee, Mo, Gonzalez, Hellerstein, Joseph, and Parameswaran}{Petersohn et~al\mbox{.}}{2020}]%
        {petersohn_towards_2020}
\bibfield{author}{\bibinfo{person}{Devin Petersohn}, \bibinfo{person}{Stephen Macke}, \bibinfo{person}{Doris Xin}, \bibinfo{person}{William Ma}, \bibinfo{person}{Doris Lee}, \bibinfo{person}{Xiangxi Mo}, \bibinfo{person}{Joseph~E. Gonzalez}, \bibinfo{person}{Joseph~M. Hellerstein}, \bibinfo{person}{Anthony~D. Joseph}, {and} \bibinfo{person}{Aditya Parameswaran}.} \bibinfo{year}{2020}\natexlab{}.
\newblock \showarticletitle{Towards scalable dataframe systems}.
\newblock \bibinfo{journal}{\emph{Proceedings of the VLDB Endowment}} \bibinfo{volume}{13}, \bibinfo{number}{12} (\bibinfo{date}{aug} \bibinfo{year}{2020}), \bibinfo{pages}{2033--2046}.
\newblock
\showISSN{2150-8097}
\urldef\tempurl%
\url{https://doi.org/10.14778/3407790.3407807}
\showDOI{\tempurl}


\bibitem[\protect\citeauthoryear{Poms, Crichton, Hanrahan, and Fatahalian}{Poms et~al\mbox{.}}{2018}]%
        {poms_scanner_2018}
\bibfield{author}{\bibinfo{person}{Alex Poms}, \bibinfo{person}{Will Crichton}, \bibinfo{person}{Pat Hanrahan}, {and} \bibinfo{person}{Kayvon Fatahalian}.} \bibinfo{year}{2018}\natexlab{}.
\newblock \showarticletitle{Scanner: efficient video analysis at scale}.
\newblock \bibinfo{journal}{\emph{ACM Transactions on Graphics}} \bibinfo{volume}{37}, \bibinfo{number}{4} (\bibinfo{date}{aug} \bibinfo{year}{2018}), \bibinfo{pages}{1--13}.
\newblock
\showISSN{0730-0301, 1557-7368}
\urldef\tempurl%
\url{https://doi.org/10.1145/3197517.3201394}
\showDOI{\tempurl}


\bibitem[\protect\citeauthoryear{Project}{Project}{2024}]%
        {gstreamer}
\bibfield{author}{\bibinfo{person}{The~GStreamer Project}.} \bibinfo{year}{2024}\natexlab{}.
\newblock \bibinfo{title}{GStreamer: open source multimedia framework}.
\newblock \bibinfo{howpublished}{\url{https://gstreamer.freedesktop.org/}}.
\newblock


\bibitem[\protect\citeauthoryear{Redmon, Divvala, Girshick, and Farhadi}{Redmon et~al\mbox{.}}{2016}]%
        {7780460}
\bibfield{author}{\bibinfo{person}{Joseph Redmon}, \bibinfo{person}{Santosh Divvala}, \bibinfo{person}{Ross Girshick}, {and} \bibinfo{person}{Ali Farhadi}.} \bibinfo{year}{2016}\natexlab{}.
\newblock \showarticletitle{You Only Look Once: Unified, Real-Time Object Detection}. In \bibinfo{booktitle}{\emph{2016 IEEE Conference on Computer Vision and Pattern Recognition (CVPR)}}. \bibinfo{pages}{779--788}.
\newblock
\urldef\tempurl%
\url{https://doi.org/10.1109/CVPR.2016.91}
\showDOI{\tempurl}


\bibitem[\protect\citeauthoryear{Roboflow}{Roboflow}{[n.d.]}]%
        {Roboflow_Supervision}
\bibfield{author}{\bibinfo{person}{Roboflow}.} \bibinfo{year}{[n.d.]}\natexlab{}.
\newblock \bibinfo{booktitle}{\emph{{Supervision}}}.
\newblock
\urldef\tempurl%
\url{https://github.com/roboflow/supervision}
\showURL{%
\tempurl}


\bibitem[\protect\citeauthoryear{Serrano}{Serrano}{2007}]%
        {serrano_programming_2007}
\bibfield{author}{\bibinfo{person}{Manuel Serrano}.} \bibinfo{year}{2007}\natexlab{}.
\newblock \showarticletitle{Programming web multimedia applications with hop}. In \bibinfo{booktitle}{\emph{Proceedings of the 15th {ACM} international conference on {Multimedia}}} \emph{(\bibinfo{series}{{MM} '07})}. \bibinfo{publisher}{Association for Computing Machinery}, \bibinfo{address}{New York, NY, USA}, \bibinfo{pages}{1001--1004}.
\newblock
\showISBNx{978-1-59593-702-5}
\urldef\tempurl%
\url{https://doi.org/10.1145/1291233.1291450}
\showDOI{\tempurl}


\bibitem[\protect\citeauthoryear{Winecki and Nandi}{Winecki and Nandi}{2024}]%
        {winecki2024_v2v}
\bibfield{author}{\bibinfo{person}{Dominik Winecki} {and} \bibinfo{person}{Arnab Nandi}.} \bibinfo{year}{2024}\natexlab{}.
\newblock \showarticletitle{{V2V}: Efficiently Synthesizing Video Results for Video Queries}. In \bibinfo{booktitle}{\emph{2024 IEEE 40th International Conference on Data Engineering (ICDE)}}. \bibinfo{pages}{5614--5621}.
\newblock
\urldef\tempurl%
\url{https://doi.org/10.1109/ICDE60146.2024.00449}
\showDOI{\tempurl}


\bibitem[\protect\citeauthoryear{Yviquel, Lorence, Jerbi, Cocherel, Sanchez, and Raulet}{Yviquel et~al\mbox{.}}{2013}]%
        {yviquel_orcc_2013}
\bibfield{author}{\bibinfo{person}{Herve Yviquel}, \bibinfo{person}{Antoine Lorence}, \bibinfo{person}{Khaled Jerbi}, \bibinfo{person}{Gildas Cocherel}, \bibinfo{person}{Alexandre Sanchez}, {and} \bibinfo{person}{Mickael Raulet}.} \bibinfo{year}{2013}\natexlab{}.
\newblock \showarticletitle{Orcc: multimedia development made easy}. In \bibinfo{booktitle}{\emph{Proceedings of the 21st {ACM} international conference on {Multimedia}}} \emph{(\bibinfo{series}{{MM} '13})}. \bibinfo{publisher}{Association for Computing Machinery}, \bibinfo{address}{New York, NY, USA}, \bibinfo{pages}{863--866}.
\newblock
\showISBNx{978-1-4503-2404-5}
\urldef\tempurl%
\url{https://doi.org/10.1145/2502081.2502231}
\showDOI{\tempurl}


\bibitem[\protect\citeauthoryear{Zaharia, Xin, Wendell, Das, Armbrust, Dave, Meng, Rosen, Venkataraman, Franklin, Ghodsi, Gonzalez, Shenker, and Stoica}{Zaharia et~al\mbox{.}}{2016}]%
        {10.1145/2934664}
\bibfield{author}{\bibinfo{person}{Matei Zaharia}, \bibinfo{person}{Reynold~S. Xin}, \bibinfo{person}{Patrick Wendell}, \bibinfo{person}{Tathagata Das}, \bibinfo{person}{Michael Armbrust}, \bibinfo{person}{Ankur Dave}, \bibinfo{person}{Xiangrui Meng}, \bibinfo{person}{Josh Rosen}, \bibinfo{person}{Shivaram Venkataraman}, \bibinfo{person}{Michael~J. Franklin}, \bibinfo{person}{Ali Ghodsi}, \bibinfo{person}{Joseph Gonzalez}, \bibinfo{person}{Scott Shenker}, {and} \bibinfo{person}{Ion Stoica}.} \bibinfo{year}{2016}\natexlab{}.
\newblock \showarticletitle{Apache Spark: a unified engine for big data processing}.
\newblock \bibinfo{journal}{\emph{Commun. ACM}} \bibinfo{volume}{59}, \bibinfo{number}{11} (\bibinfo{date}{oct} \bibinfo{year}{2016}), \bibinfo{pages}{56–65}.
\newblock
\showISSN{0001-0782}
\urldef\tempurl%
\url{https://doi.org/10.1145/2934664}
\showDOI{\tempurl}


\bibitem[\protect\citeauthoryear{Zhang, Daum, He, Balazinska, Haynes, and Krishna}{Zhang et~al\mbox{.}}{2023}]%
        {zhang_equi-vocal_2023}
\bibfield{author}{\bibinfo{person}{Enhao Zhang}, \bibinfo{person}{Maureen Daum}, \bibinfo{person}{Dong He}, \bibinfo{person}{Magdalena Balazinska}, \bibinfo{person}{Brandon Haynes}, {and} \bibinfo{person}{Ranjay Krishna}.} \bibinfo{year}{2023}\natexlab{}.
\newblock \showarticletitle{{EQUI}-{VOCAL}: {Synthesizing} {Queries} for {Compositional} {Video} {Events} from {Limited} {User} {Interactions} [{Technical} {Report}]}.
\newblock \bibinfo{journal}{\emph{arXiv preprint arXiv:2301.00929}} (\bibinfo{year}{2023}).
\newblock
\urldef\tempurl%
\url{https://doi.org/10.48550/arXiv.2301.00929}
\showDOI{\tempurl}


\bibitem[\protect\citeauthoryear{Zhang, Yuan, Liu, Sun, Liang, Jin, and Wang}{Zhang et~al\mbox{.}}{2019}]%
        {zhang_fast_2019}
\bibfield{author}{\bibinfo{person}{Pengju Zhang}, \bibinfo{person}{Chunmiao Yuan}, \bibinfo{person}{Kunliang Liu}, \bibinfo{person}{Yukuan Sun}, \bibinfo{person}{Jiayu Liang}, \bibinfo{person}{Guanghao Jin}, {and} \bibinfo{person}{Jianming Wang}.} \bibinfo{year}{2019}\natexlab{}.
\newblock \showarticletitle{Fast {Video} {Clip} {Retrieval} {Method} via {Language} {Query}}. In \bibinfo{booktitle}{\emph{Advanced {Data} {Mining} and {Applications}}}, \bibfield{editor}{\bibinfo{person}{Jianxin Li}, \bibinfo{person}{Sen Wang}, \bibinfo{person}{Shaowen Qin}, \bibinfo{person}{Xue Li}, {and} \bibinfo{person}{Shuliang Wang}} (Eds.). \bibinfo{publisher}{Springer International Publishing}, \bibinfo{address}{Cham}, \bibinfo{pages}{526--534}.
\newblock
\showISBNx{978-3-030-35231-8}
\urldef\tempurl%
\url{https://doi.org/10.1007/978-3-030-35231-8_38}
\showDOI{\tempurl}


\bibitem[\protect\citeauthoryear{Zhang, Li, Han, and Wen}{Zhang et~al\mbox{.}}{2022}]%
        {zhang_ai_2022}
\bibfield{author}{\bibinfo{person}{Xinrong Zhang}, \bibinfo{person}{Yanghao Li}, \bibinfo{person}{Yuxing Han}, {and} \bibinfo{person}{Jiangtao Wen}.} \bibinfo{year}{2022}\natexlab{}.
\newblock \showarticletitle{{AI} {Video} {Editing}: a {Survey}}.
\newblock \bibinfo{journal}{\emph{Preprints}} (\bibinfo{date}{feb} \bibinfo{year}{2022}).
\newblock
\urldef\tempurl%
\url{https://doi.org/10.20944/preprints202201.0016.v2}
\showDOI{\tempurl}
\newblock
\shownote{Publisher: Preprints.}


\end{thebibliography}

\end{document}